\documentclass[a4paper,11pt]{article}
\pdfoutput=1 

\usepackage{jheppub} 

\usepackage[T1]{fontenc} 
\usepackage{besphysics}
\usepackage{booktabs}
\usepackage{multirow}
\usepackage{subfigure}
\usepackage{rotating}
\usepackage{lineno}
\usepackage{hyperref}

\newcommand{\kl}{K^0_L}

\newcommand{\klpipi}{\kl\pi^+\pi^-}
\newcommand{\kslpipi}{K^0_{S,L}\pi^+\pi^-}

\newcommand{\kmpipipi}{K^-\pi^+\pi^+\pi^-}

\newcommand{\kmppipipi}{K^\mp\pi^\pm\pi^+\pi^-}
\newcommand{\kpmpipipi}{K^\pm\pi^\mp\pi^+\pi^-}
\newcommand{\kmppipio}{K^\mp\pi^\pm\pi^0}
\newcommand{\kpmpipio}{K^\pm\pi^\mp\pi^0}
\newcommand{\kmppi}{K^\mp\pi^\pm}
\newcommand{\kpmpi}{K^\pm\pi^\mp}

\newcommand{\BESIIIorcid}[1]{\href{https://orcid.org/#1}{\hspace*{0.1em}\raisebox{-0.45ex}{\includegraphics[width=1em]{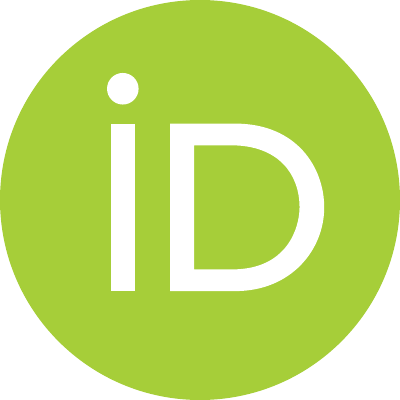}}}}

\title{\boldmath Improved measurements of the coherence factors and strong-phase differences in $D\to K^-\pi^+\pi^+\pi^-$ and $D\to K^-\pi^+\pi^0$ with quantum-correlated $D\bar{D}$ decays}

\collaboration{BESIII Collaboration}

\author{
M.~Ablikim$^{1}$\BESIIIorcid{0000-0002-3935-619X},
M.~N.~Achasov$^{4,b}$\BESIIIorcid{0000-0002-9400-8622},
P.~Adlarson$^{78}$\BESIIIorcid{0000-0001-6280-3851},
X.~C.~Ai$^{83}$\BESIIIorcid{0000-0003-3856-2415},
R.~Aliberti$^{37}$\BESIIIorcid{0000-0003-3500-4012},
A.~Amoroso$^{77A,77C}$\BESIIIorcid{0000-0002-3095-8610},
Q.~An$^{74,60,\dagger}$,
Y.~Bai$^{59}$\BESIIIorcid{0000-0001-6593-5665},
O.~Bakina$^{38}$\BESIIIorcid{0009-0005-0719-7461},
Y.~Ban$^{48,g}$\BESIIIorcid{0000-0002-1912-0374}
% ...
\\
\vspace{0.2cm}
(BESIII Collaboration)\\
\vspace{0.2cm} {\it
$^{1}$ Institute of High Energy Physics, Beijing 100049, People's Republic of China\\
% ...
$^{4}$ Budker Institute of Nuclear Physics SB RAS (BINP), Novosibirsk 630090, Russia\\
% ...
$^{37}$ Johannes Gutenberg University of Mainz, Johann-Joachim-Becher-Weg 45, D-55099 Mainz, Germany\\
$^{38}$ Joint Institute for Nuclear Research, 141980 Dubna, Moscow region, Russia\\
% ...
$^{48}$ Peking University, Beijing 100871, People's Republic of China\\
% ...
$^{59}$ Southeast University, Nanjing 211100, People's Republic of China\\
$^{60}$ State Key Laboratory of Particle Detection and Electronics, Beijing 100049, Hefei 230026, People's Republic of China\\
% ...
$^{74}$ University of Science and Technology of China, Hefei 230026, People's Republic of China\\
% ...
$^{77}$ University of Turin and INFN, (A)University of Turin, I-10125, Turin, Italy; (B)University of Eastern Piedmont, I-15121, Alessandria, Italy; (C)INFN, I-10125, Turin, Italy\\
$^{78}$ Uppsala University, Box 516, SE-75120 Uppsala, Sweden\\
% ...
$^{83}$ Zhengzhou University, Zhengzhou 450001, People's Republic of China\\

\vspace{0.2cm}
$^{\dagger}$ Deceased\\
% ...
$^{b}$ Also at the Novosibirsk State University, Novosibirsk, 630090, Russia\\
% ...
$^{g}$ Also at State Key Laboratory of Nuclear Physics and Technology, Peking University, Beijing 100871, People's Republic of China\\
% ...

}
}

\abstract{Improved measurements of the coherence factors and strong-phase differences in $D\to K^-\pi^+\pi^+\pi^-$ and $D\to K^-\pi^+\pi^0$ decays are reported, using quantum-correlated $D\bar{D}$ pairs produced in $\ee$ annihilation at a center-of-mass energy of $3.773\gev$, where $D$ denotes a quantum superposition of the flavour-specific $\Dz$ and $\Dzbar$ mesons. The analysis employs a dataset collected by the BESIII experiment, corresponding to an integrated luminosity of $7.93~\rm fb^{-1}$. The observables sensitive to the coherence factors and strong-phase differences are measured by reconstructing one $D$ meson in the signal mode and the other in a tag mode.
These parameters provide essential inputs to the measurement of the angle $\gamma$ of the Cabibbo-Kobayashi-Maskawa Unitarity Triangle in the LHCb and Belle II experiments. The coherence factors are determined to be $R_{K3\pi}=0.51\pm0.04$ and $R_{K\pi\pi^0}=0.75\pm0.03$, and the strong-phase differences are $\delta_D^{K3\pi}=\left(182^{+14}_{-13}\right)^\circ$ and $\delta_D^{K\pi\pi^0}=\left(209^{+7}_{-8}\right)^\circ$, where the uncertainties include both statistical and systematic contributions.  For $D\to\kmpipipi$, the parameters have been further determined in four phase-space bins with improved precision compared to the previous BESIII results. The uncertainty on future $\gamma$ measurements from the knowledge of $D\to\kmpipipi$ parameters is expected to be reduced to approximately 3.5$^\circ$.}

\begin{document} 
\maketitle
\flushbottom

\section{Introduction}
\label{sec:intro}
$C\!P$ violation in the Standard Model arises from the irreducible complex phase of the Cabibbo-Kobayashi-Maskawa (CKM) quark-mixing matrix~\cite{Cabibbo,Kobayashi:1973fv}.
A precise determination of the angle $\gamma=\arg(-V_{us}{V_{ub}}^*/V_{cs}{V_{cb}}^*)$, which can be measured in the tree-level decay $B^- \to D K^-$ with negligible theoretical uncertainty, is an important goal of heavy flavour physics. Throughout this paper, $D$ denotes a superposition of $\Dz$ and $\Dzbar$ mesons reconstructed in a final state common to both. 
The decays $D \to K^- n \pi$ ($n \geq 1$) are an important category of charm decays~\cite{ADS0,ADS}.
They can proceed via Cabibbo-favoured (CF) amplitudes, doubly Cabibbo-suppressed (DCS) amplitudes, or through their interference. 
Among these, the final states $D\to\kpmpipipi$ and $D\to\kpmpipio$ are important for the determination of $\gamma$. Their sensitivity to $\gamma$ in analyses of $B \to Dh$ decays depends on knowledge of the hadronic parameters:
the coherence factor $R_S$, the amplitude ratio $r_D^{S}$, and the $C\!P$-conserving strong-phase difference $\delta_D^{S}$ between the CF and DCS amplitudes~\cite{Atwood:2003mj}. These parameters can be defined as 
\begin{equation}
R_{S}e^{-i\delta_{D}^{S}}=\frac{\int {\cal A}^{\star}_{S}(\textbf{x}){\cal A}_{\bar{S}}(\textbf{x})\rm d\textbf{x}}{A_{S}A_{\bar{S}}} {\;\;\;{\rm and}\;\;\;} 
r^{S}_{D} = A_{\bar{S}} / A_{S}.
\label{eq:rds}
\end{equation}
Here, ${\cal A}_{S}(\textbf{x})$ represents the decay amplitude of $\Dz \to S$ at the point $\textbf{x}$ in the multi-body phase space, and $A^2_{S}=\int |{\cal A}_{S}(\textbf{x})|^2 {\rm d}\textbf{x}$, with an analogous expression for $\bar{S}$. The coherence factor, which lies between 0 and 1, quantifies the degree of interference and is influenced by the resonance structures of the final state.
The amplitude ratio parameter $r_D^{S}$, which characterizes the relative strength of the DCS to CF processes, is typically of order $\lambda^2 \approx 0.05$, where $\lambda$ is the Wolfenstein parameter defined in Ref.~\cite{Wolfenstein:1983yz}.

The coherence factor and strong-phase difference can be extracted using $D\bar{D}$ pairs produced in $e^+e^-$ collisions at the $\psi(3770)$ resonance by studying both the $D$ and $\bar{D}$ mesons decays in each event. The joint decay rate depends on the hadronic parameters. 
This approach has been applied to the datasets of the CLEO-c experiment~\cite{Lowery:2009id,Libby:2014rea,Evans:2016tlp} and the BESIII experiment~\cite{Harnew:2013wea}. The recent BESIII measurement yielded the results for the decay \mbox{$D \to \kmthreepi$}, \mbox{$R_{K3\pi} = 0.52^{+0.12}_{-0.10}$}, $\delta_D^{K3\pi}=\left(167^{+31}_{-19}\right)^\circ$ and $r_D^{K3\pi} = (5.46 \pm 0.09) \times 10^{-2}$; and for the decay $D \to \kmpipio$, $R_{K\pi\pi^0} = 0.78 \pm 0.04$, $\delta_D^{K\pi\pi^0}=\left(196^{+14}_{-15}\right)^\circ$ and $r_D^{K\pi\pi^0} = (4.40 \pm 0.11) \times 10^{-2}$~\cite{BESIII:2021eud}, where the uncertainties include both statistical and systematic contributions. Complementary constraints on these parameters can also be derived from measurements of $D^0$-$\bar{D}^{0}$ oscillations above charm threshold~\cite{Harnew:2013wea,LHCb:2025zgk}. 

It has been found that partitioning the phase space of the decay $D\to \kmthreepi$ into regions enhances its utility as one of the most sensitive channels for determining the angle $\gamma$~\cite{Evans:2019wza}. 
This improvement arises from the strong-phase variation across the phase space and the presence of regions with high coherence. 
A binning scheme to exploit this feature has been proposed, based on amplitude models of the decays $D^0 \to \kmthreepi$ and $D^0 \to \kpthreepi$ developed by LHCb~\cite{Aaij:2017kbo}. 
Interpreting the $B$ decay rates in these bins requires knowledge of the corresponding binned information for the $D$ hadronic parameters.
The proof-of-principle study for measuring the binned parameters was first performed with the CLEO-c dataset~\cite{Evans:2019wza}. Subsequently, BESIII performed the measurements with $\ee$ annihilation data at a center-of-mass energy of
 $\sqrt{s}=3.773\gev$ corresponding to an integrated luminosity of 2.93 $\ifb$ collected in 2010 and 2011~\cite{BESIII:2021eud}. 
Taking the binned parameters as inputs, the $C\!P$ asymmetries and related observables in the $B^- \to DK^-$ decays have been measured by LHCb~\cite{LHCb:2022nng}, highlighting that the uncertainties from the input charm hadronic parameters now contribute at a level comparable to statistical uncertainties. Further refinement of these inputs is critical to reduce this dominant systematic uncertainty.

This paper presents the improved measurements of the coherence factors and average strong-phase differences for $D \to \kmthreepi$ and $D \to \kmpipio$, obtained by integrating over the full phase space each mode. The results from a binned phase-space analysis for $D \to \kmthreepi$ are also reported. The analysis is based on a dataset corresponding to an integrated luminosity of $7.93\,{\rm fb^{-1}}$ (including that used in a previous BESIII measurement~\cite{BESIII:2021eud}), which is collected by the BESIII experiment in $e^+e^-$ collisions at $\sqrt{s}=3.773\gev$ during 2010, 2011 and 2022~\cite{Ablikim:2014gna,Ablikim:2015orh}. Compared with the previous BESIII measurement, this analysis also benefits from the inclusion of a new $D\to\klpipi$ tag mode, and a more refined background treatment. These improvements lead to a significant enhancement in the precision of the measured hadronic parameters.
The paper is structured as follows: Section~\ref{sec:formalism} defines the measured observables and their connection to the underlying physics parameters. A description of the BESIII detector and the Monte Carlo simulations used in the analysis is provided in Section~\ref{sec:detector}. The event selection criteria are detailed in Section~\ref{sec:selection}. Section~\ref{sec:obs} presents the measured observables, the methodology for extracting hadronic parameters, and the evaluation of systematic uncertainties. The impact of the results on measurements of $\gamma$ in the $B^- \to DK^-$ decays is discussed in Section~\ref{sec:gamma}, and the paper concludes with a summary in Section~\ref{sec:summary}.

\section{Formalism and measurement strategy}
\label{sec:formalism}

The neutral $D\bar{D}$ pairs from $\ee$ collisions at the $\psi(3770)$ resonance are quantum-correlated and are produced with no additional final-state particles. The two mesons are in a $\mathcal{C}$-odd eigenstate, which implies an antisymmetric wave function for the $\Dz\Dzbar$ system.

Both the $\Dz$ and $\Dzbar$ decays into specific final states, denoted as $S$(signal) and $T$(tag).  An event is referred to as {\it double-tag} when both $S$ and $T$ are reconstructed. Three distinct classes of tag modes are employed in this analysis: $C\!P$ tags, defined as the decays into $C\!P$ eigenstates (or states close to them); like-sign tags, identified by a charged kaon in the tag decay having the same sign as that in the signal decay; and self-conjugate $\kslpipi$ tags. These are first described in the context of the global measurement, in which all events across the signal decay phase space are treated uniformly, and then discussed for the case where the $D\to \kmthreepi$ decay is analyzed using a binned approach. 

In the case where the tagging mode is a $C\!P$ eigenstate, the decay rate can be expressed as
\begin{equation}
\Gamma(S|C\!P) =  A_{S}^{2}A_{C\!P}^{2}\,  \left( 1 + (r_D^S)^{2}  -  2 \lambda R_S r_D^S  \cos{\delta_D^{S}} \right),
\label{eq:cptag}
\end{equation}
where $C\!P$ indicates a specific mode with eigenvalue $\lambda=+1(-1)$ and $\delta_D^T=0(\pi)$ for pure $C\!P$-even(-odd) tags. Some self-conjugate decay modes are not pure $C\!P$ eigenstates but are predominantly $C\!P$ even or $C\!P$ odd. In this case, Eq.~(\ref{eq:cptag}) remains valid with $\lambda = (2F_+^T - 1)$, where $F_+^T$ is the $C\!P$-even fraction of the decay mode. The decay $D\to\pipipio$ is a well-known case with $F_+^{\pi\pi\pi^0}>95\%$~\cite{Nayak:2014tea,Malde:2015mha,BESIII:2024nnf}.

It is useful to define the observables $\rho^S_{C\!P\pm}$, which represent the ratios of the number of $C\!P$-tagged signal events to the number expected in the absence of quantum correlations:
\begin{equation}
\rho_{C\!P\pm}^{S} \equiv \frac{ N(S|C\!P)+N(\bar{S}|C\!P)} 
{2N_{D\bar{D}}\,\br{}\left( D^{0}\to C\!P \right) \left[\br{}\left(D^{0}\to S\right)+\br{}\left( D^{0}\to \bar{S}\right) \right]}. 
\label{eq:rho_cp}
\end{equation}
Here, $N(S|C\!P)$ 
($N(\bar{S}|C\!P)$) denotes the efficiency-corrected yield of decays to the channel $S$ ($\bar{S}$) tagged with a $C\!P$ eigenstate, $N_{D\bar{D}}$ is the number of neutral $D\bar{D}$ pairs produced in the sample, and $\br{}(D^0 \to S/\bar{S})$ is the branching fraction of the $D^0$ meson to the final state $S/\bar{S}$. The observable $\rho^S_{C\!P+}$ ($\rho^S_{C\!P-}$) corresponds to the case when the $C\!P$ tag is even (odd). 

It is necessary to account for corrections arising from $D^0$-$\bar{D}^{0}$ oscillations when relating the squared amplitudes to their branching fractions.
Retaining terms up to ${\cal{O}}\left(r_D^2, x^2,y^2 \right)$, the following relations apply:
\begin{eqnarray}
CF:&\br{}(D^0\to S) =& A^2_S\left(1 - R_Sr^S_D(y\cos\delta^S_D + x\sin\delta^S_D) + (y^2-x^2)/2 \right), \label{eq:BRCF}\\
DCS:&\br{}(D^0\to \bar{S}) =& A^2_{{S}}\left((r_D^S)^2 - R_S r_D^S (y \cos \delta^S_D - x \sin\delta^S_D) + (x^2 + y^2)/2 \right), \label{eq:BRDCS}\\ 
CP:&\br{}(D^0 \to C\!P) =& A^2_{C\!P}\left(1 - \lambda y + y^2 \right)\,. \label{eq:BRCP}
\end{eqnarray}
Note that both the charm mixing parameters $x$ and $y$ are of order $\sim r^S_D/10$~\cite{LHCb-CONF-2024-004}. 
The observable $\rho^S_{C\!P\pm}$ can then be expressed in terms of the $D$ decay parameters as:
\begin{equation}
\begin{aligned}
\rho _ {C\!P\pm} ^ { S } = 
\frac{ 1 + (r_{D}^{S})^2 - 2\lambda r_{D}^{S} R_{S} \cos \delta_{D}^{S}}{1  -\lambda y + (r_D^S)^2 -2r_D^SR_Sy\cos\delta^S_D}.
\end{aligned}
\label{eq:rho_cp_the}
\end{equation}
This and subsequent expressions are written to ${\cal{O}}\left(r_D^2, x,y \right)$.

In practice, the precision of the $\rho_{C\!P\pm}^S$ measurement with many $C\!P$ eigenstates is limited by the uncertainties on the branching fractions of the tagging modes. This limitation can be largely mitigated by re-expressing $\rho_{C\!P}^S$ in terms of $N(K^\mp\pi^\pm|C\!P)$, the efficiency-corrected yield of the $D\to K^\mp\pi^\pm$ decays tagged by the same $C\!P$ eigenstate. Furthermore, to remove the influence of imperfect simulation of reconstruction efficiencies, the branching fractions of $D^0\to K^\mp\pi^\pm$ and $D^0\to S/\bar{S}$ can be replaced by the efficiency-corrected single-tag yields $N_{ST}$, where only one $D$ decaying to the final states $K^\mp\pi^\pm$ or $S,\bar{S}$ is selected. Then, from Eq.~(\ref{eq:rho_cp}), it follows:
\begin{equation}
\frac{\rho_{C\!P\pm}^S}{\rho_{C\!P\pm}^{K\pi}}=\frac{N(S|C\!P)+N(\bar{S}|C\!P)}{N(K^-\pi^+|C\!P)+N(K^+\pi^-|C\!P)} \,
\frac{N_{ST}(D\to K^-\pi^+)}{N_{ST}(D\to S)} \, .
\label{eq:kpinorm}
\end{equation}
The values of $\rho_{C\!P\pm}^{K\pi}$ can be calculated  from Eq.~(\ref{eq:rho_cp_the}) with a precision of $0.5\%$.

Consider the case where the tagging mode is also a flavoured $K^- n \pi$ final state, for example $K^-\pi^+$, $\kmthreepi$ or $\kmpipio$, and the kaon charge in the tagging mode matches that in the signal mode. When the signal and tag modes are identical, the corresponding decay rate is given by:
\begin{equation}
\Gamma(S|S) =  A_{S}^{2}A_{\bar{S}}^{2}\,  (1 - R_S^2).
\label{eq:lstag}
\end{equation}
An observable is then defined as the ratio of the number of efficiency-corrected double-tagged events, $N(S|S)$ and $N(\bar{S}|\bar{S})$, to the expected yield in the absence of quantum correlations:
\begin{equation}
\rho^S_{LS}  \equiv  \frac{N(S|S)+N(\bar{S}|\bar{S})}{2N_{D\bar{D}}\,\br{}( D^{0}\to S)\br{}( D^{0}\to \bar{S})}. \label{eq:rho_ls} 
\end{equation}
Since quantum-correlation effects are negligible in opposite-sign double tags, it is experimentally advantageous to evaluate this observable using the equivalent expression:
\begin{equation}
\rho^S_{LS} \, \frac{\br{}(D^0 \to \bar{S})}{\br{}(D^0 \to S)} = \frac{N(S|S)+N(\bar{S}|\bar{S})}{2N(S|\bar{S})}. \label{eq:rho_ls2}
\end{equation}
Expressing $\rho^S_{LS}$ in terms of the $D$ decay parameters leads to
\begin{equation}
\rho _ { L S } ^ { S } =
\frac{  1 -  R _ { S }  ^ { 2 } }{ 1  - R_{S} \left((y/r_D^S)\cos\delta_{D}^{S}-(x/r_D^S)\sin\delta_{D}^{S}\right) + (x^2 + y^2)/(2(r_D^S)^2)},
\label{eq:rho_ls2_th}
\end{equation}
from which it is apparent that this observable is highly sensitive to the coherence factor.

When the tag is the two-body mode $D \to K^-\pi^+$, the coherence factor is $R_T = 1$, and the parameters $r_D^{K\pi}$ and $\delta_D^{K\pi}$ are well-determined from the measurements of $D^0$-$\bar{D}^{0}$ oscillations. The case $S=\kmthreepi$ and $T=\kmpipio$ simultaneously provides sensitivity to both multi-body signal modes of interest.
The quantum-correlated decay rate is given by:
\begin{equation}
\Gamma(S|T) =  A_{S}^{2}A_{T}^{2}\,  [(r_D^S)^{2} + (r_D^T)^2 -  2R_S R_T r_D^S r_D^T \cos{(\delta_D^{T}-\delta_D^{S})} ],
\label{eq:lstagdiff}
\end{equation}
where $T \ne S$. The quantum-correlation effects in these double-tagged events can subsequently be studied through the observables
\begin{equation}
\rho^S_{T,LS} \equiv
\frac{ N(S|T)+N(\bar{S}|\bar{T})} 
{2N_{D\bar{D}}\,\left[ \br{}( D^{0}\to S) \, \br{}(D^{0}\to \bar{T})+
	\br{}( D^{0}\to \bar{S}) \, \br{}(D^{0}\to {T}) \right]}.
\label{eq:rho_lsx}
\end{equation}
As before, it is advantageous to exploit the yields of opposite-sign double tags and to evaluate an equivalent expression
\begin{equation}
\rho^S_{T,LS} \, \left(\frac{\br{}(D^0 \to \bar{T})}{\br{}(D^0 \to T)} +\frac{\br{}(D^0 \to \bar{S})}{\br{}(D^0 \to S)}\right)=
\frac{ N(S|T)+N(\bar{S}|\bar{T})}{N(S|\bar{T})+N(\bar{S}|T)}.
\label{eq:rho_lsx2}
\end{equation}
It follows from Eqs.~(\ref{eq:lstagdiff}), (\ref{eq:rho_lsx}) and the branching fraction relations that
\begin{equation}
\rho_{T,LS}^S=\frac{(r_D^S)^2+(r_D^{T})^2-2R_SR_Tr_D^Sr_D^{T}\cos\left(\delta_D^{T}-\delta_D^S\right)}{(r_D^S)^2+(r_D^{T})^2-R_Sr_D^S\left(y\cos\delta_D^S-x\sin\delta_D^S\right)-R_Tr_D^{T}\left(y\cos\delta_D^{T}-x\sin\delta_D^{T}\right)+x^2+y^2}.
\end{equation}

For the self-conjugate tag $D\to \kslpipi$, the Dalitz plots can be divided into eight pairs of bins symmetric about the line $m_-^2 = m_+^2$, where $m_-^2 = m(\ks\pi^-)^2$ and $m_+^2= m(\ks \pi^+)^2$. The binning scheme is developed based on an amplitude model of the decay $D\to \kspipi$~\cite{Aubert:2008bd}. The bin boundaries are chosen such that each bin spans an equal range of the strong-phase difference, defined as \mbox{$\Delta \delta_D^{{\nkspipi}}\equiv \delta_D^{\nkspipi}(m_+^2,m_-^2) - \delta_D^{\nkspipi}(m_-^2,m_+^2)$}. This strategy, referred to as the ``equal-$\Delta \delta_D$ binning scheme'', is illustrated in Figure~\ref{fig:kspipidalitz}.

\begin{figure}
	\centering
	\includegraphics[width=.48\textwidth]{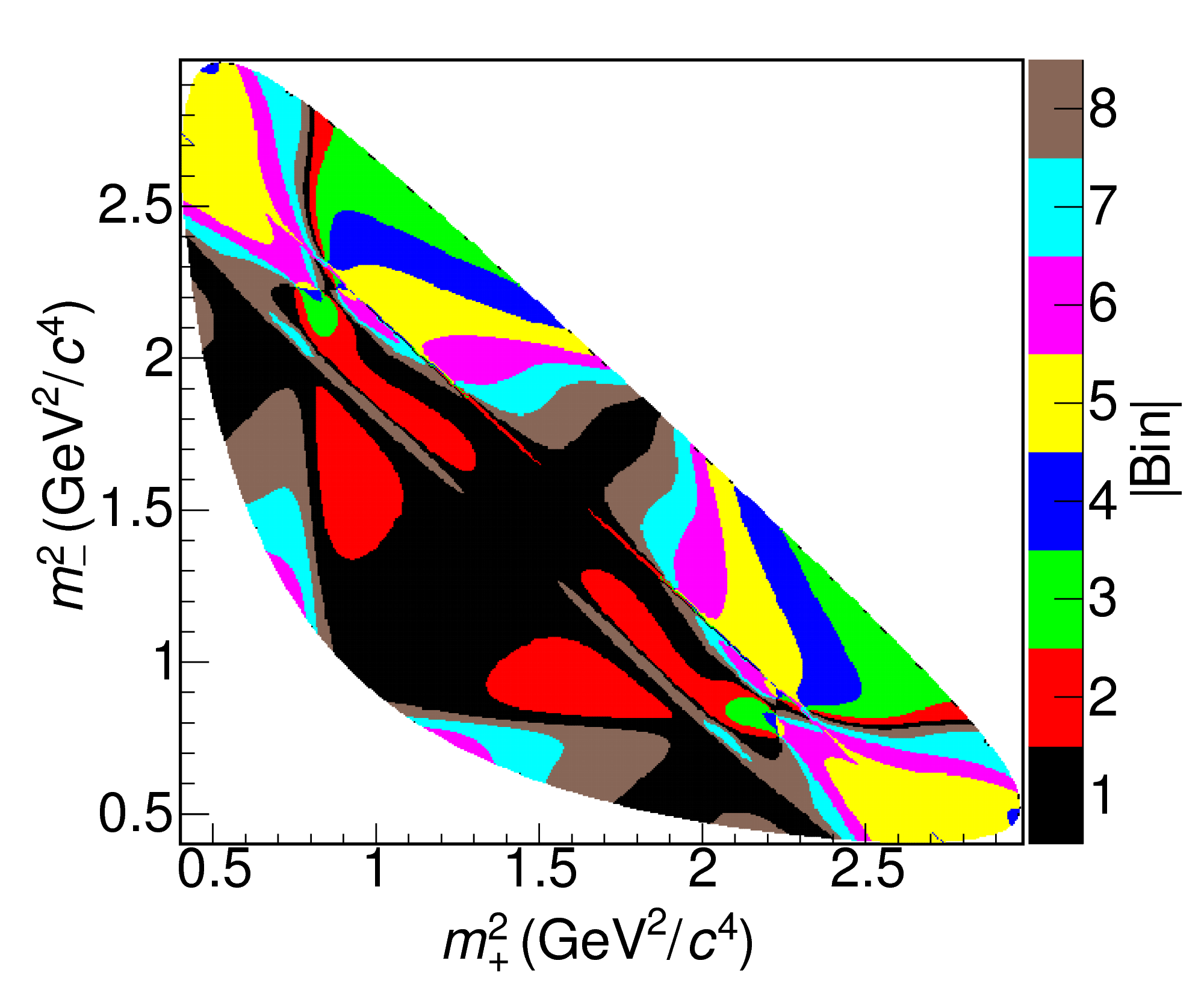}
	\caption{The Dalitz plot of $D \to \kspipi$ illustrating the equal-$\Delta \delta_D$ binning scheme.}
	\label{fig:kspipidalitz}
\end{figure}

Measurements performed with quantum-correlated $D\bar{D}$ pairs determine the parameters $c_i$ and $s_i$, which correspond to the amplitude-weighted averages of the cosine and sine, respectively, of the strong-phase difference in the bin $i$ of the $D\to K^0_{S}\pi^+\pi^-$ Dalitz plot:
\begin{align}
c_i &\equiv \frac{\int_{i} {\rm d}{m_+^2} \, {\rm d}{m_-^2} \, |{\cal{A}}_{K^0_{S}\pi\pi}({m_+^2},{m_-^2})| |{\cal{A}}_{K^0_{S}\pi\pi}({m_-^2},{m_+^2})| \cos
	\Delta \delta_D^{K^0_{S}\pi\pi}({m_+^2},{m_-^2})}
{\sqrt{\int_{i} {\rm d}{m_+^2} \, {\rm d}{m_-^2} \, |{\cal{A}}_{K^0_{S}\pi\pi}({m_+^2},{m_-^2})|^2 \int_{i} {\rm d}{m_+^2} \, {\rm d}{m_-^2} \, |{\cal{A}}_{K^0_{S}\pi\pi}({m_-^2},{m_+^2})|^2}}\,,
\label{eq:ci}
\end{align}
with an analogous expression for $s_i$. The strong-phase difference parameters for the decay $D \to \klpipi$, denoted $c_i^{\prime}$ and $s_i^{\prime}$, are defined similarly to $c_i$ and $s_i$ with the same binning scheme as shown in Figure~\ref{fig:kspipidalitz}.

When employing $D\to K^0_{S(L)}\pi^+\pi^-$ as a tag mode, it is also necessary to know $K_i^{(\prime)}$, which is the probability of a flavour-specific $D^0$ decay occurring in the bin $i$:
\begin{equation}
\label{eq:ki}
K_i^{(\prime)} = \frac{\int_{i} {\rm d}{m_+^2} {\rm d}{m_-^2} |{\cal A}_{K^0_{S(L)}\pi\pi}(m_+^2,m_-^2)|^2 }{\int {\rm d}{m_+^2} {\rm d}{m_-^2} |{\cal A}_{K^0_{S(L)}\pi\pi}(m_+^2,m_-^2  )|^2},
\end{equation}
where the denominator is the sum of the integrals over all bins.
This quantity can be measured in flavour-tagged decays, either at charm threshold or at higher energies (in the latter case, small corrections must be applied to account for the effects of $D^0$-$\bar{D}^{0}$ oscillations~\cite{Bondar:2010qs}). The decay modes $D\to \kslpipi$ have been extensively studied at charm threshold, and the measurements of the $c_i^{(\prime)}$, $s_i^{(\prime)}$ and $K_i^{(\prime)}$ parameters have been performed by both the CLEO and BESIII Collaborations~\cite{Libby:2010nu,Ablikim:2020yif,Ablikim:2020lpk,BESIII:2025nsp}.

Events in which one meson decays to $K^- n\pi$ ($K^+ n\pi$) and the other to $K_{S,L}^0\pi^+\pi^-$ are assigned a negative bin number if $m_-^2<m_+^2$ ($m_-^2>m_+^2$).
With these definitions and assuming uniform acceptance over the Dalitz plot, the event yield is given by:
\begin{equation}\label{eq_kspipi}
\begin{aligned}
Y^{S}_{i} =  H \left( K _ { i } + \left( r _ { D } ^ {S} \right) ^ { 2 } K _ { - i } -   2 r _ { D } ^ {S} R_S \sqrt { K _ { i } K _ { - i } } \left[ c _ { i } \cos \delta _ { D } ^ {S} - s _ { i } \sin \delta _ { D } ^ {S} \right] \right)
\end{aligned}
\end{equation}
for $D\to\kspipi$ and
\begin{equation}\label{eq_klpipi}
\begin{aligned}
{Y^{\prime}}^{ S}_{i} =  H^{\prime} \left( K _ { i }^{\prime} + \left( r _ { D } ^ {S} \right) ^ { 2 } K _ { - i }^{\prime} +   2 r _ { D } ^ {S} R_S \sqrt { K _ { i }^{\prime} K _ { - i }^{\prime} } \left[ c _ { i }^{\prime} \cos \delta _ { D } ^ {S} - s _ { i }^{\prime} \sin \delta _ { D } ^ {S} \right] \right)
\end{aligned}
\end{equation}
for $D\to\klpipi$, where $H^{(\prime)}$ is a bin-independent normalisation factor.

In Ref.~\cite{Evans:2019wza} it is shown that the sensitivity to $\gamma$ in $B^- \to DK^-$ with $D\to \kmthreepi$ can be improved by dividing the $D$ decay phase space into bins, each characterized by its own coherence factor and strong-phase difference. A binning scheme is proposed based on the resonance substructure of the CF and DCS decays, as modelled in the LHCb study~\cite{Aaij:2017kbo}. The phase space is divided into four bins, each spanning a range of the strong-phase difference between the $D^0$ and $\bar{D}^{0}$ decay amplitudes. The bin boundaries are chosen to equalise the product of the integrated CF and DCS amplitudes between bins, which is a metric shown to optimise the $\gamma$ sensitivity. A small region of phase space, where the invariant mass of the $\pi^+\pi^-$ pair lies close to the mass of the $\ks$ meson, is excluded to suppress the background from $D \to \ks K^-\pi^+$. According to the amplitude models, this exclusion removes about $5\%$ of the signal. Since the phase space of $D\to\kmthreepi$ is five-dimensional, there is no convenient way to visualise the bins. However, each decay is unambiguously assigned to a bin using the four-momenta of the final-state particles.

The observables can be determined within each bin in the same way as in the global analysis, yielding four measurements each for $\rho^{S}_{C\!P\pm}$, $\rho^{S}_{F,LS}$ and $Y^{S}_i$. For $\rho^{K3\pi}_{LS}$, where both the signal and tag decays are $D\to \kmthreepi$, ten distinct measurements are obtained.

\section{Detector and simulation samples}
\label{sec:detector}

The BESIII detector~\cite{Ablikim:2009aa} records symmetric $e^+e^-$ collisions 
produced by the BEPCII storage ring~\cite{Yu:IPAC2016-TUYA01}
in the center-of-mass energy range from 1.84 to 4.95~GeV,
with a peak luminosity of $1.1 \times 10^{33}\;\text{cm}^{-2}\text{s}^{-1}$ 
achieved at $\sqrt{s} = 3.773\;\text{GeV}$. 
BESIII has collected large data samples in this energy region~\cite{Ablikim:2019hff,Lu:2020imt,Zhang:2022bdc}. The cylindrical core of the BESIII detector covers 93\% of the full solid angle and consists of a helium-based
multilayer drift chamber~(MDC), a plastic scintillator time-of-flight
system~(TOF), and a CsI(Tl) electromagnetic calorimeter~(EMC),
which are all enclosed in a superconducting solenoidal magnet
providing a 1.0~T magnetic field.
The solenoid is supported by an
octagonal flux-return yoke with resistive plate counter muon
identification modules interleaved with steel. 
The charged-particle momentum resolution at $1~{\rm GeV}/c$ is
$0.5\%$, and the 
${\rm d}E/{\rm d}x$
resolution is $6\%$ for electrons
from Bhabha scattering. The EMC measures photon energies with a
resolution of $2.5\%$ ($5\%$) at $1$~GeV in the barrel (end cap)
region. The time resolution in the TOF barrel region is 68~ps, while
that in the end cap region was 110~ps. The end cap TOF
system was upgraded in 2015 using multigap resistive plate chamber
technology, providing a time resolution of
60~ps,
which benefits 63\% of the data used in this analysis~\cite{BESIII:2009fln}.

Simulated samples produced with the {\sc
	geant4}-based~\cite{geant4} Monte Carlo  package, which
includes the geometric description of the BESIII detector and the
detector response, are used to determine the detection efficiencies
and to estimate the backgrounds. The simulation includes the beam-energy spread of 0.97\,MeV and initial-state radiation (ISR) in the $e^+e^-$
annihilations modelled with the generator {\sc
	kkmc}~\cite{ref:kkmc}. The inclusive Monte Carlo samples consist of the production of $D^0\bar{D}^0$ and $D^+D^-$ pairs from decays of the $\psi(3770)$,
decays of the $\psi(3770)$ to charmonia or light hadrons, the ISR
production of the $J/\psi$ and $\psi(3686)$ states, and the
continuum processes incorporated in {\sc kkmc}~\cite{ref:kkmc}. The equivalent integrated luminosity of the inclusive Monte Carlo samples is about forty times that of the data. The known decay modes are modelled with {\sc
	evtgen}~\cite{Lange:2001uf,Rong_Gang_2008}
using branching fractions taken from the
Particle Data Group~\cite{ParticleDataGroup:2024cfk}, and the remaining unknown decays
from the charmonium states with {\sc
	lundcharm}~\cite{PhysRevD.62.034003,
	YANGRui-Ling:61301}. The final-state radiation from charged final-state particles is incorporated with the {\sc
	photos} package~\cite{RICHTERWAS1993163}. The signal processes are generated separately taking the spin-matrix elements into account in {\sc evtgen}. 
The signal mode $D \to \kmpipio$ is generated with a resonance substructure which matches that of the CF process, and the DCS process is modeled with a phase-space distribution; for the decay $D \to \kmthreepi$ the simulated events are generated by the CF and DCS models developed by LHCb~\cite{LHCb:2017swu}. 
Sample sizes of 1,500,000 events are simulated for each class of double tag, apart from the case where the signal mode is $D\to \kmpi$, for which 750,000 events are produced for each signal channel. The $D\to K^\mp K_S^0\pi^\pm$ processes are also generated with phase-space model and weighted using Dalitz plot from Ref.~\cite{LHCb:2015lnk} to estimate the $D\to K^\mp K_S^0\pi^\pm$ background in $D \to \kmthreepi$.
The quantum-correlation effects have been incorporated into the simulated samples where applicable.

\section{Event selection and yield determination}
\label{sec:selection}

\subsection{Selection of double-tagged events}

To determine the observables introduced in Section~\ref{sec:formalism}, double-tagged samples are used, in which one charm meson is reconstructed in the signal mode, and the other in a tag mode classified as either flavour, $C\!P$, or self-conjugate tag.  
Flavour tags are classified as like-sign, where the kaon has the same charge as that in the signal mode, and opposite-sign, which are used for normalisations.
Six $C\!P$-even tags are reconstructed, including the quasi-eigenstate mode $D\to \pipipio$, together with six $C\!P$-odd tags.
Events in which one charm meson decays to $\kmpi$ and the other to a $C\!P$ tag mode are also reconstructed for normalisations.
The self-conjugate category includes both $D \to \kspipi$ and $D \to \klpipi$ tag modes. The complete list of tags used in this analysis is provided in Table~\ref{tab:taglist}.
Events with a $\kl$ meson in the final state are reconstructed using a missing-mass technique, as described below, while all other double-tagged events are fully reconstructed. Effects of $C\!P$ violation and matter-interaction effects in the neutral-kaon system are negligible compared with the experimental sensitivity~\cite{Bjorn:2019kov}. 
Following the selection criteria described in Ref.~\cite{BESIII:2021eud}, charged kaon and pion tracks, as well as photons, are reconstructed and combined to form the mesons that decay within the detector.
These mesons are reconstructed through the modes: $\ks \to \pi^+\pi^-$, $\pi^0\to\gamma\gamma$, $\eta\to\gamma\gamma$ and $\pi^+\pi^-\pi^0$, $\omega\to\pi^+\pi^-\pi^0$, $\eta'\to\pi^+\pi^-\eta$ and $\gamma\pi^+\pi^-$, and $\phi \to K^+K^-$. In like-sign tags, in order to suppress the misidentified backgrounds from the opposite-sign mode, the track is labelled a $K$ ($\pi$) candidate if the probabilities $P_K>100 \,  P_\pi$ ($P_\pi>100 \, P_K$), where these probabilities are determined by combining the d$E$/d$x$ measurement in the MDC and the time-of-flight information from the TOF under the $K$ and $\pi$ hypotheses.

\begin{table}[!ht]
	\caption{The tag modes used in conjunction with the signal decays $D \to \kmthreepi$ and $D \to \kmpipio$.  The $C\!P$ tags are also reconstructed against the decay $D \to \kmpi$.  Charge-conjugate modes are always implied}
	\label{tab:taglist}
	\begin{center}
		\begin{tabular}{llccc}
			\toprule
			\multirow{2}{*}{Flavour} & Like sign  &$\kmthreepi$, $\kmpipio$, $\kmpi$\\
			& Opposite sign  &$\kpthreepi$, $\kppipio$, $\kppi$\\
			\multirow{2}{*}{$C\!P$} & Even &$\kk$, $\pipi$, $\kspiopio$, $\klpio$, $\klomega$, $\pipipio$\\
			& Odd &$\kspio$, $\kseta$, $\ksomega$, $\ksetap$, $\ksphi$, $\klpiopio$\\
			\multicolumn{2}{l}{Self-conjugate}  & $\kspipi$, $\klpipi$\\
			\bottomrule
		\end{tabular}
	\end{center}
\end{table}

Following Ref.~\cite{BESIII:2021eud}, to suppress combinatorial background, the energy difference $\Delta{E} = E_{D} - \sqrt{s}/2$ is required to satisfy the selection criteria. A $\ks$ veto based on flight distance is applied to $D \to \pipipio$ candidates to suppress the background from $D \to \kspio$, and similarly to
$D\to \kmthreepi$ candidates to reduce the contamination from $D \to \ks K^-\pi^+$. 
If multiple double-tagged candidates are found in a single event, the candidate with the average of the reconstructed invariant masses closest to the nominal $D^0$ mass is retained. Approximately 10\% of selected events contain more than one candidate.

For each fully-reconstructed $D$ candidate, the beam-constrained mass $M_{\rm BC}$ is calculated to provide optimal separation between signal and combinatorial background:
\begin{equation}
M_{\rm BC} = \sqrt{(\sqrt{s}/2)^2/c^4-|{\mathbf{p_{D}}}|^2/c^2},
\label{eq:mbc}
\end{equation}
where $\mathbf{p_{D}}$ is the momentum of the $D$ candidate in the rest frame of the initial $e^+e^-$ collision.   Figures~\ref{fig:mbcexamples1} and~\ref{fig:mbcexamples2} show the $M_{\rm BC}$ distributions of each signal decay for a selection of double tags.

\begin{figure}
	\centering
	\includegraphics[width=.99\textwidth]{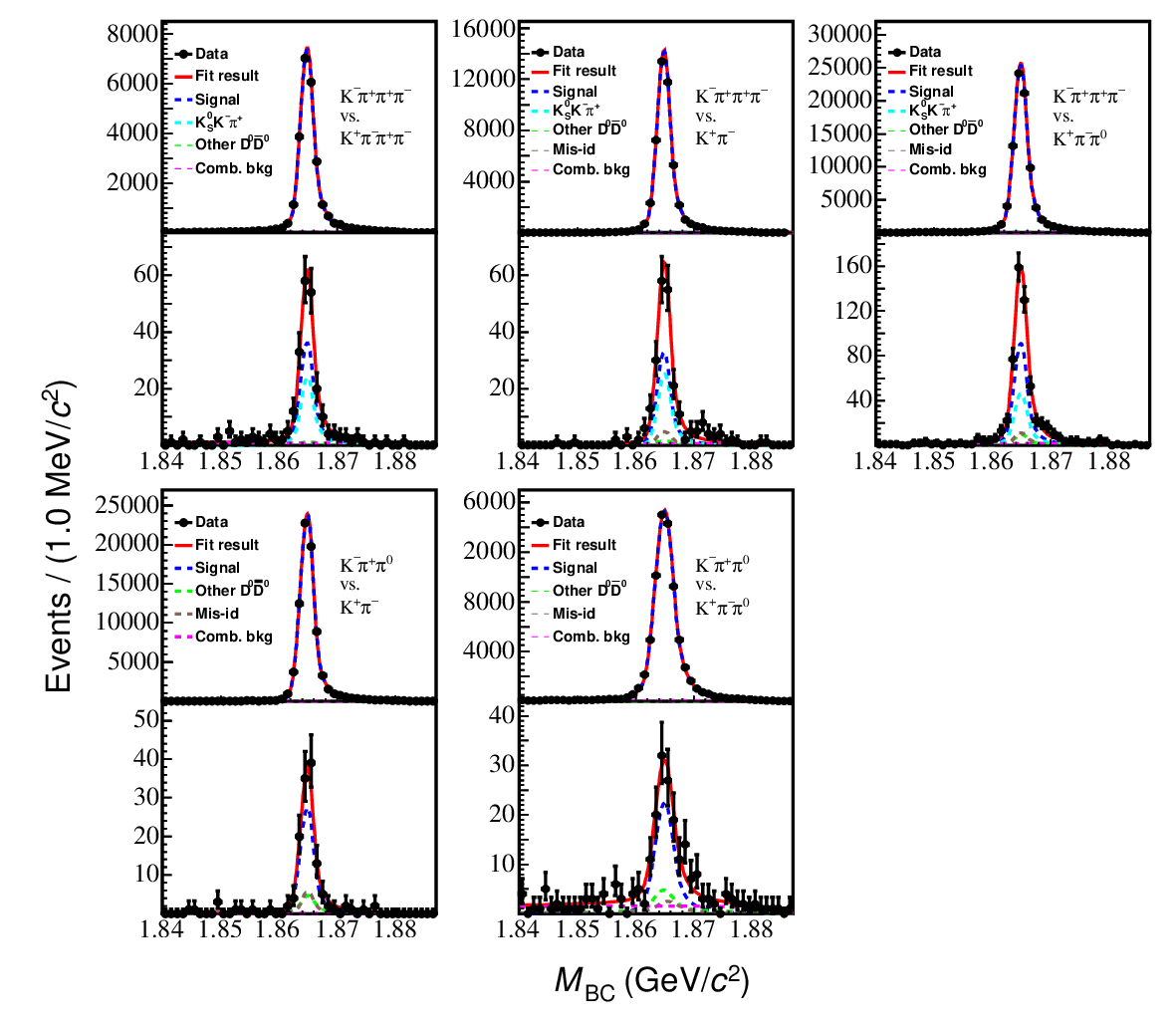}
    \put(-165,70){LS}
    \put(-290,70){LS}
    \put(-40,240){LS}
    \put(-165,240){LS}
    \put(-290,240){LS}
    \put(-165,135){OS}
    \put(-290,135){OS}
    \put(-40,305){OS}
    \put(-165,305){OS}
    \put(-290,305){OS}
	\caption{The $M_{\rm BC}$ distributions for the like-sign and opposite-sign flavour tags. The data are shown as points with error bars. The red lines represent the total fit; the long-dashed cyan lines indicate the $D \to K_S^0 K^\mp \pi^\pm$ background; the misidentified opposite-sign background in like-sign case is shown as the dashed browm lines, and other $D^0\bar{D}^0$ backgrounds are shown as the dashed green lines. The dashed magenta lines represent the combinatorial background, which is very small.}
	\label{fig:mbcexamples1}
\end{figure}

\begin{figure}
	\centering
	\includegraphics[width=.99\textwidth]{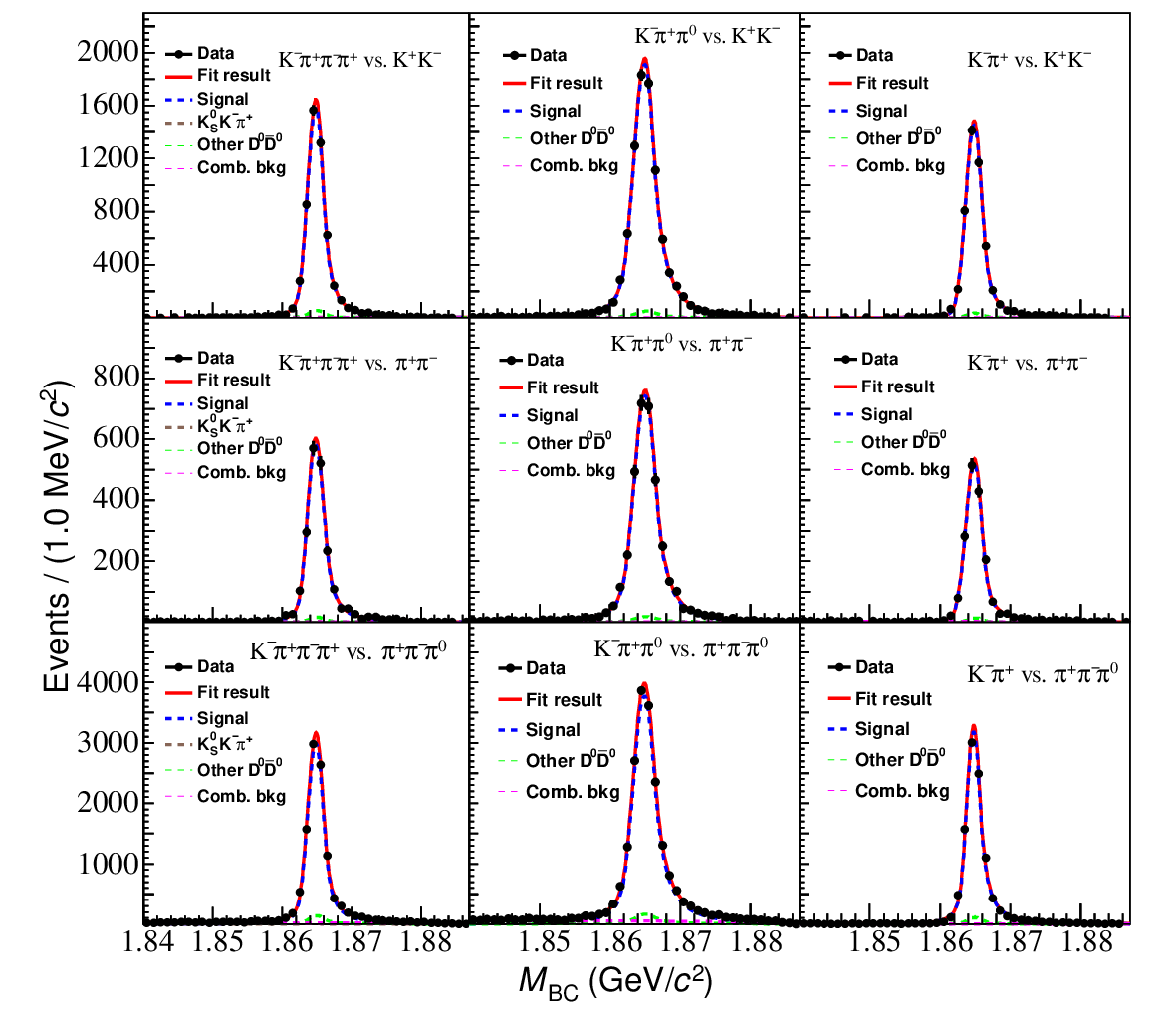}
	\caption{The $M_{\rm BC}$ distributions for a selection of $C\!P$-tagged decays.
		The points with error bars are data; the red lines indicate the total fit; the dashed brown and green lines show the peaking-background contributions from $D\to K_S^0 K^\mp \pi^\pm$ and other $D^{0}\bar{D}^{0}$ process, respectively; the combinatorial-background contributions are shown as the dashed magenta line.}
	\label{fig:mbcexamples2}
\end{figure}

Double-tagged events where the tag side involves a $K^0_L$ meson cannot be fully reconstructed.  As described in Ref.~\cite{BESIII:2021eud}, these events are selected using a missing-mass technique.  The missing-mass squared is calculated as
\begin{equation}
M_{\rm miss}^2 = (\sqrt{s}/2-E_{X})^2/c^4-|\mathbf{p_{S}}+\mathbf{p_{X}}|^2/c^2,
\label{eq:mm2}
\end{equation}
where $\mathbf{p_S}$ denotes momentum of the signal mode, while $E_{X}$ and $\mathbf {p_{X}}$ denote the total energy and momentum, respectively, of the charged particles and $\pi^0$ candidates that are not associated with the signal mode.
It is expected to peak at the squared mass of the $K^0_{L}$ meson for the $C\!P$ tags under consideration. To suppress contamination from $\ks \to \pi^0\pi^0$ and other backgrounds, events are rejected if they contain extra $\pi^0$ candidates, extra charged tracks, any $\eta \to \gamma \gamma$ candidates, or multiple $\pi^0$ candidates that share common showers.  Figure~\ref{fig:mmexamples} shows the $M_{\rm miss}^2$ distributions for double-tagged events containing a $K^0_{L}$ meson. To suppress the $D\to\ks\pi^{0}(\ks\to\pi^{0}\pi^{0})$ and $D\to\kl\pi^{0}$ background in the $D\to\kl\pi^{0}\pi^{0}$ tag mode, an extra requirement $M^{2}_{\rm miss}(\pi^{0})>0.49\gev^2/c^4$ is applied for each $\pi^{0}$ in $D\to\kl\pi^{0}\pi^{0}$.

\begin{figure}
	\centering
	\includegraphics[width=.99\textwidth]{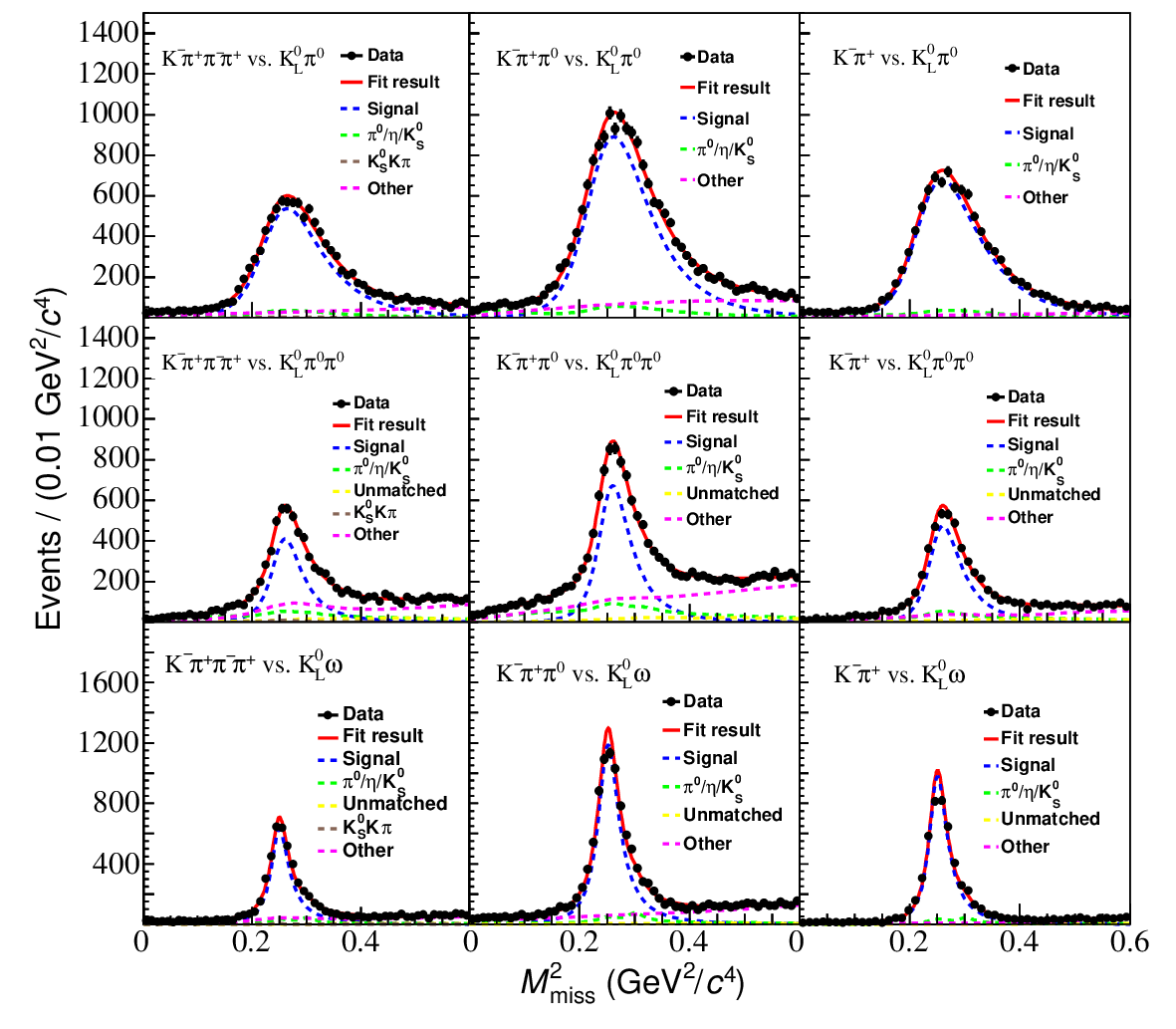}
	\caption{The $M_{\rm miss}^2$ distributions for double tags involving a $K^0_L$ meson. The points with error bars represent the data; the red lines denote the total fit; the dashed blue lines indicate the signal component; the dashed brown lines correspond the $D\to K_S^0 K^\mp \pi^\pm$ background; dashed green curves represent the combined $\pi^0 X$, $\eta X$, and $\ks X$ peaking backgrounds; and the dashed yellow and magenta curves represent the unmatched and other $D^{0}\bar{D}^0$ background, respectively.}
	\label{fig:mmexamples}
\end{figure}

\subsection{Signal yield determination and consideration of peaking backgrounds}
\label{sec:bckpeak}

The signal yield for the double tags is determined from an extended unbinned maximum-likelihood fit to the $M_{\rm BC}$ distribution of each signal decay for fully reconstructed events, and to the $M_{\rm miss}^2$ distribution for the events containing a $K^0_L$ candidate. The signal shape is modelled using fits to the simulated samples, employing a Crystal Ball function for $M_{\rm BC}$ and a Johnson's $S_U$ function~\cite{Johnson:1949} for $M_{\rm miss}^2$. The shape parameters are fixed according to Monte Carlo simulation, except for those describing the central value and width, which are allowed to vary to account for the resolution difference between data and simulation. The combinatorial background in the $M_{\rm BC}$ distribution is described with an ARGUS function~\cite{ALBRECHT1990278}, while that in the $M_{\rm miss}^2$ distribution is modeled using a {\sc RooKeysPdf}~\cite{Cranmer:2000du} determined from simulation.
In general, the contributions from peaking backgrounds within the signal region are fixed based on Monte Carlo simulation, with their shapes described by the {\sc RooKeysPdf}. For fully reconstructed events, these peaking backgrounds are typically small, contributing less than 5\%, 10\%, and 10\% in the flavour-tagged, $C\!P$-tagged, and $D\to \kspipi$-tagged samples, respectively. However, certain backgrounds are notably larger or require dedicated treatment, as discussed in detail below. All fits are observed to converge with satisfactory residual distributions.

The dominant peaking backgrounds in the tag modes containing a $K_L^0$ arise from the decay channels involving an $\eta$, $\pi^0$ or $K_S^0$, where these particles are missing and thus contribute to the $M^2_{\rm miss}$ spectrum. The shapes and contributions of these backgrounds are fixed based on Monte-Carlo simulation.

The residual peaking backgrounds are studied and evaluated following the procedure as described in Ref.~\cite{BESIII:2021eud}, for  both the global and binned analyses of all the tag modes. Quantum-correlation effects are corrected in the simulation, where necessary, using either the decay amplitudes, the hadronic parameters, or the $C\!P$ content.

Another significant source of background in the like-sign samples arises from opposite-sign events of the same flavour tag in which both a kaon and pion of opposite charge are misidentified. To quantify the level of this contamination after applying the tight particle-identification requirement, the misidentification rates are determined in bins of momentum using a control sample of opposite-sign events selected without particle-identification requirements on one of the kaon or pion candidates.
This sample includes all flavour double-tag categories listed in Table~\ref{tab:taglist}.
The results observed in data are consistent with those from simulation, as shown in Figure~\ref{fig:misid}. Those differences are applied as corrections in the estimation of misidentified background.
This background is found to be negligible in the $D\to K^-\pi^+\pi^+\pi^-$ channels, while it contributes approximately 3\% per side in the $D\to K^-\pi^+\pi^0$ channels and about 7\% per side in the $D\to K^-\pi^+$ channels.

\begin{figure}
	\centering
	\includegraphics[width=.48\textwidth]{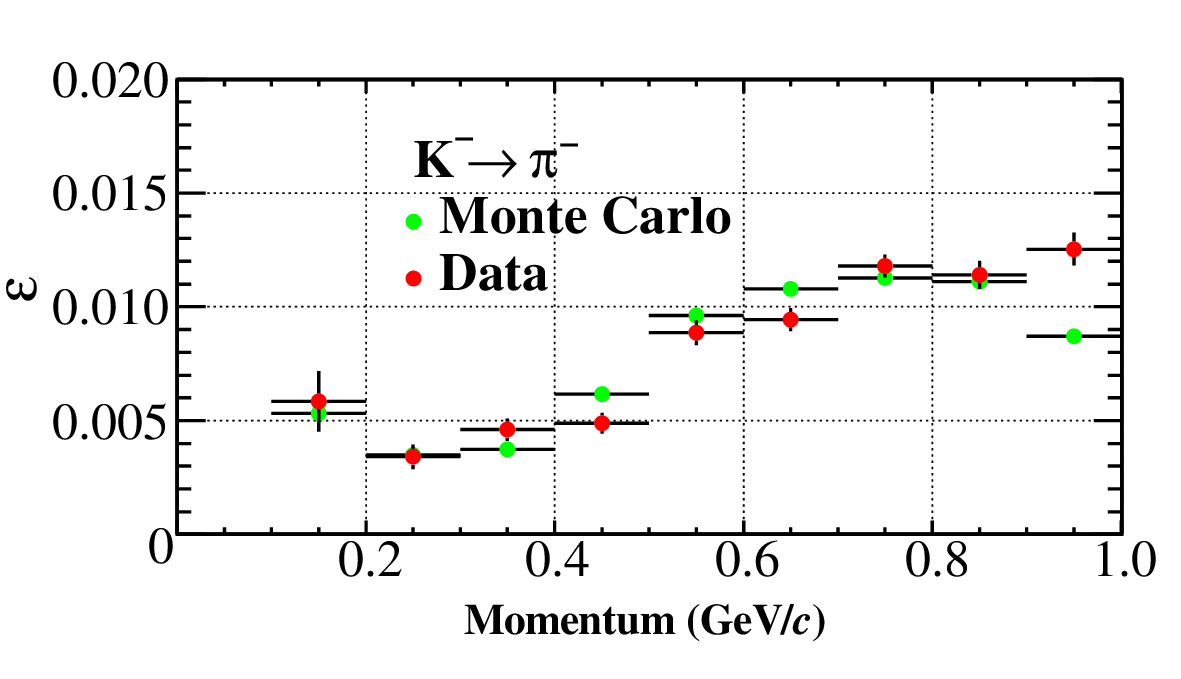}
	\includegraphics[width=.48\textwidth]{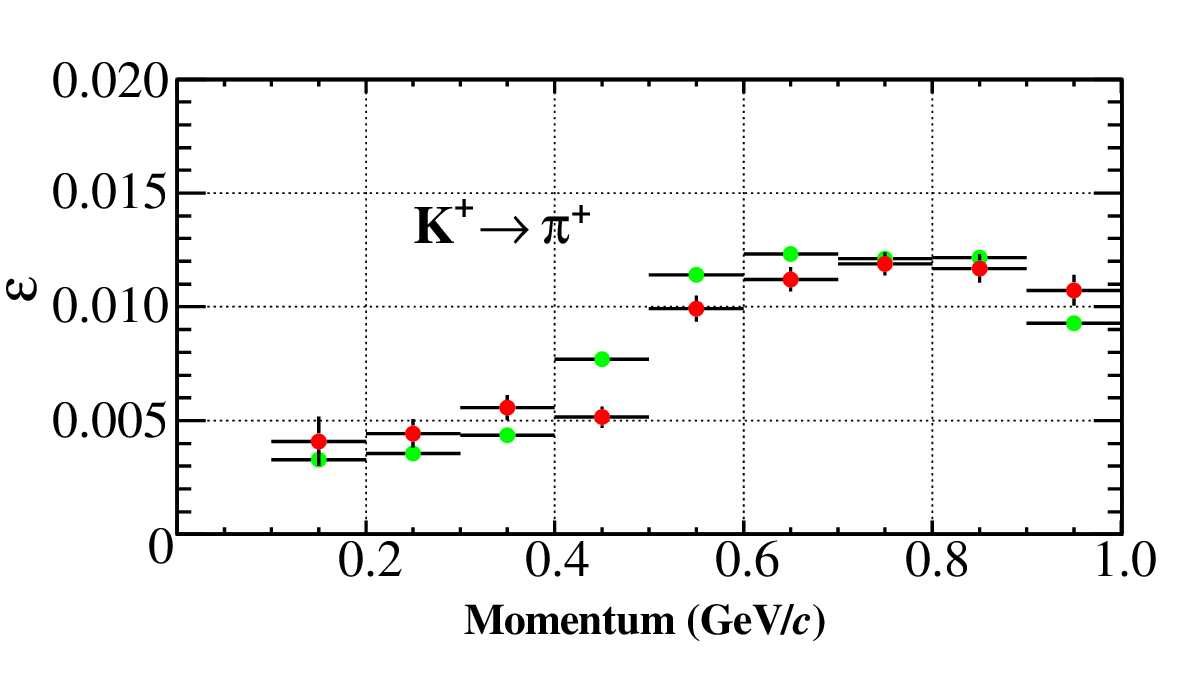}
	\includegraphics[width=.48\textwidth]{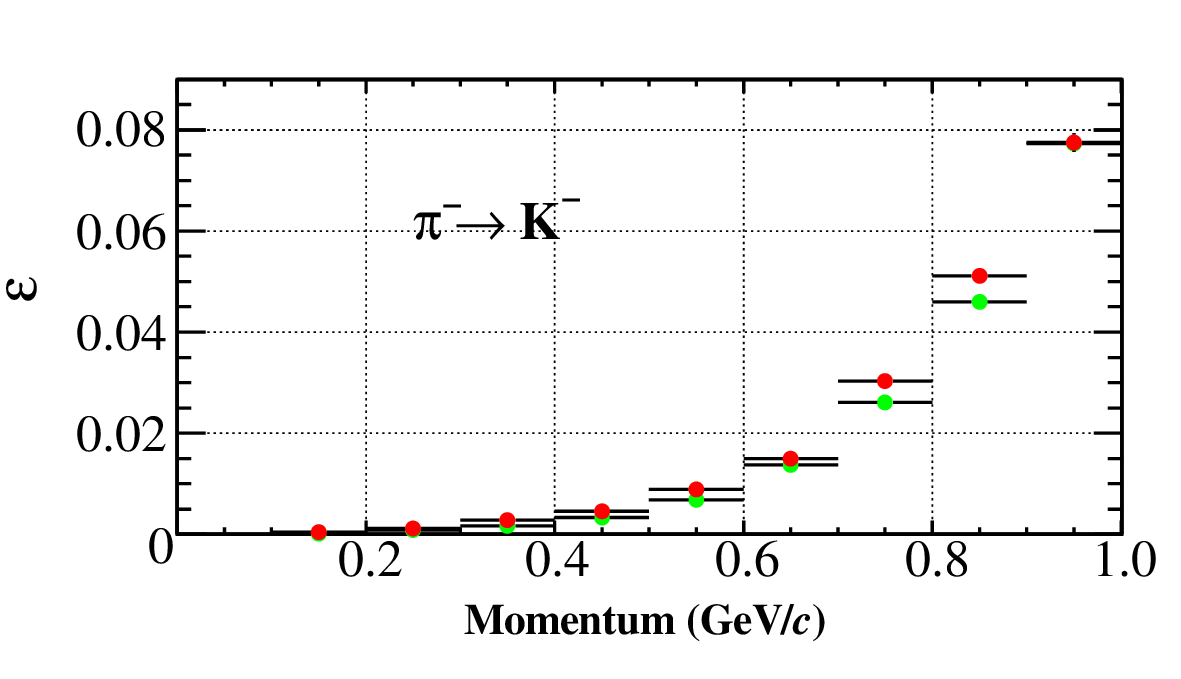}
	\includegraphics[width=.48\textwidth]{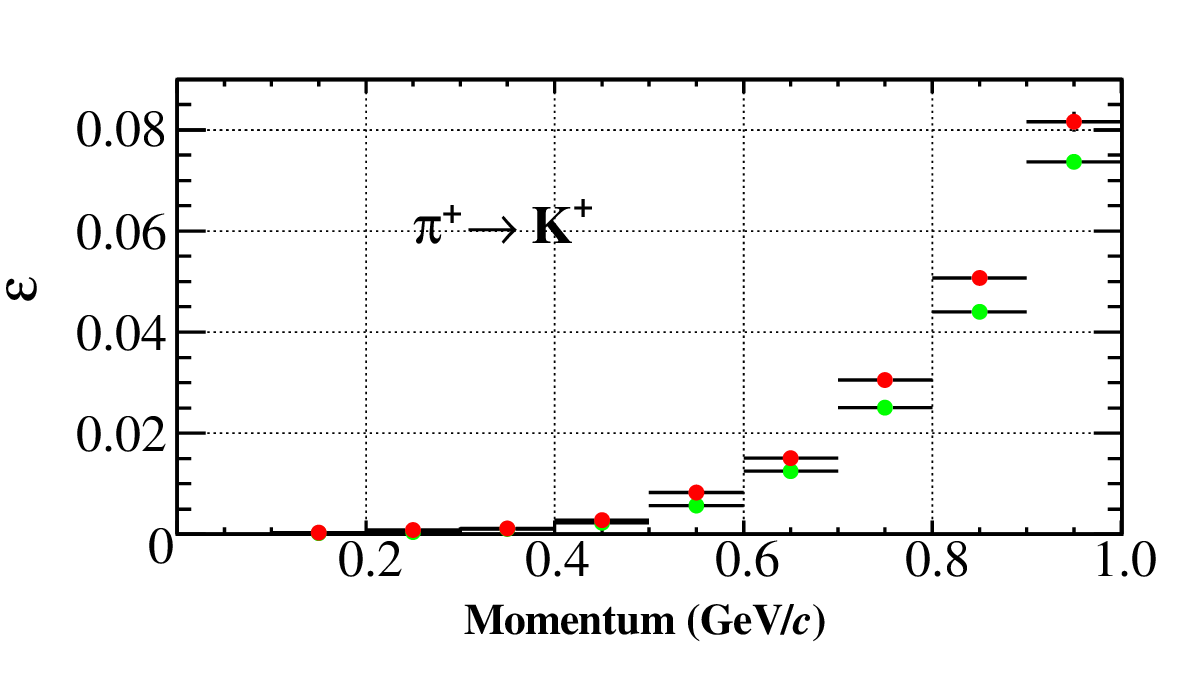}
	\caption{Efficiencies of $K \to \pi$ (top) and $\pi \to K$ (bottom) misidentifications in bins of momentum, as measured in data and Monte Carlo simulation.}
	\label{fig:misid}
\end{figure}

The signal yields for the flavour-tagged and the $D \to K_{S,L}^0\pi^+\pi^-$ tagged events are summarized in Table~\ref{tab:flavouryields}. The corresponding yields for the $C\!P$ tags are presented in Table~\ref{tab:cpyields}, together with the selection efficiencies used in determining the observables $\rho^{K3\pi}_{C\!P\pm}$ and $\rho^{K\pi\pi^0}_{C\!P\pm}$.

\begin{table}[!ht]
	\caption{Signal yields for the flavour-tagged and $D \to \kslpipi$ tagged events, where the uncertainties are statistical only. The corresponding selection efficiencies, determined from Monte Carlo simulation, are also shown; all have statistical uncertainties below $0.1\%$.}\label{tab:flavouryields}
	\begin{center}
            \begin{tabular}{lcccc}
			\toprule
Tag & \multicolumn{2}{c}{$\kmthreepi$} & \multicolumn{2}{c}{$\kmpipio$} \\
 & Yield & \multicolumn{1}{l}{$\varepsilon\ (\%)$} & Yield & $\varepsilon\ (\%)$ \\
			\midrule
$\kppi$ & 46761$\pm$226 & 25.2 & 77629$\pm$284 & 23.8 \\
$\kppipio$ & 84594$\pm$307 & 12.4 & 71100$\pm$275 & 11.6 \\
$\kpthreepi$ & 24712$\pm$167 & 12.7 & $-$ & $-$ \\
 &  &  &  &  \\
$\kmpi$ & 109.0$\pm$15.7 & 19.4 & 89.4$\pm$11.3 & 18.3 \\
$\kmpipio$ & 302.6$\pm$24.5 & 11.2 & 104.1$\pm$13.7 & 10.7 \\
$\kmthreepi$ & 120.8$\pm$15.1 & 11.6 & $-$ & $-$ \\
 &  &  &  &  \\
$\kspipi$ & 14066$\pm$128 & 14.2 & 22879$\pm$164 & 12.9 \\
$\klpipi$ & 30699$\pm$215 & 20.7 & 53521$\pm$291 & 20.6 \\
			\bottomrule
\end{tabular}
	\end{center}
\end{table}

\begin{table}[!ht]
	\caption{Signal yields for the $C\!P$-tagged events, where the uncertainties are statistical only. The corresponding selection efficiencies, determined from Monte Carlo simulation, are also shown; all have statistical uncertainties below $0.1\%$.}\label{tab:cpyields}
	\begin{center}
		\begin{tabular}{lcc cc cc} \toprule
			Tag & \multicolumn{2}{c}{$\kmthreepi$} & \multicolumn{2}{c}{$\kmpipio$} & \multicolumn{2}{c}{$\kmpi$} \\
			& Yield\hspace*{0.65cm} & $\varepsilon\ (\%)$ & Yield\hspace*{0.65cm} & $\varepsilon\ (\%)$ & Yield\hspace*{0.65cm} & $\varepsilon\ (\%)$ \\
			\midrule
$\kk$                     & $5276\pm74$  & 26.5 & $8783\pm97$   & 25.1 & $4707\pm70$  & 44.5 \\
$\pipi$                   & $2017\pm47$  & 27.9 & $3461\pm62$   & 27.3 & $1680\pm42$  & 48.1 \\
$\pipipio$                & $9956\pm110$ & 15.3 & $18247\pm161$ & 15.2 & $9794\pm102$ & 28.6 \\
$\kspiopio$               & $1990\pm50$  & 6.6  & $3617\pm72$   & 6.3  & $1990\pm50$  & 12.8 \\
$\klpio$                  & $8326\pm192$ & 19.2 & $14234\pm269$ & 17.5 & $9418\pm111$ & 37.5 \\
$\klomega$                & $3376\pm90$  & 7.8  & $6615\pm137$  & 8.1  & $4035\pm86$  & 16.5 \\
\\
$\kspio$                  & $5418\pm77$  & 16.0 & $9423\pm102$  & 15.0 & $4589\pm69$  & 29.0 \\
${\kseta}(\gamma\gamma)$  & $726\pm30$   & 13.3 & $1241\pm40$   & 12.5 & $647\pm27$   & 24.8 \\
${\kseta}(\pi\pi\pi^0)$   & $240\pm18$   & 7.7  & $415\pm26$    & 7.5  & $232\pm18$   & 14.4 \\
${\ksetap}(\gamma\pi\pi)$ & $686\pm28$   & 7.8  & $1134\pm37$   & 7.6  & $580\pm26$   & 14.3 \\
${\ksetap}(\pi\pi\eta)$   & $251\pm16$   & 5.7  & $432\pm21$    & 5.7  & $233\pm16$   & 11.1 \\
${\ksomega}$              & $1629\pm49$  & 5.7  & $2820\pm66$   & 5.5  & $1332\pm43$  & 10.8 \\
$\ksphi$                  & $299\pm27$   & 7.4  & $530\pm35$    & 7.4  & $221\pm25$   & 13.3 \\
$\klpiopio$               & $3530\pm132$ & 7.8  & $5937\pm143$  & 7.0  & $3457\pm81$  & 16.8 \\
			\bottomrule
		\end{tabular}
	\end{center}
\end{table}

To improve the resolution of the reconstructed position in phase space, a kinematic fit is applied to the $D \to \kmthreepi$ candidates, constraining their invariant masses to the known $D^0$ mass. The events are partitioned into four phase space bins defined in Ref.~\cite{Evans:2019wza}, and the same procedure described above is used to determine the signal yields in each bin. As an example, the signal yields and the statistical uncertainties of $D \to \kmthreepi$ tagged with $D \to \kspio$ are $1314 \pm 37$, $1228 \pm 34$, $1092 \pm 32$ and $1558 \pm 39$ for the bins 1 to 4, respectively.

\subsection{Efficiency-corrected double-tagged yields in $D \to \kslpipi$ bins}

The $D \to \kslpipi$ tagged events are categorized into sixteen phase-space bins of the tag decay. To improve the resolution of the decay position in phase space, a mass-constrained fit is performed, in which the $K^0_S$ mass (or the missing mass for $K^0_L$) and the $D^0$ mass are constrained. Monte Carlo studies indicate that the correct bin is identified in 91\% (85\%) of cases for the positive bins and in 95\% (91\%) for the negative bins in $D\to K_{S(L)}^0\pi^+\pi^-$.
Signal yields in each bin are determined, and efficiency corrections are then applied.
These corrections account for bin-to-bin efficiency variations, which fall within $\pm 15\%$, as well as for migration between bins.  The corrected signal yields, $Y^{K3\pi, K\pi\pi^0}_i$ and ${Y^{\prime}}^{K3\pi,K\pi\pi^0}_i$, are listed in Tables~\ref{tab:kspipiglobal} and \ref{tab:klpipiglobal}, respectively.  
The same procedure is applied to the binned $D \to \kmthreepi$ analysis to determine the signal yields. For instance, in the bin 1 of the $D \to \kspipi$ phase-space, the efficiency-corrected signal yields and their statistical uncertainties of $D \to \kmthreepi$ are $3841\pm155$, $3767\pm151$, $4149\pm159$, and $5003\pm184$ for the bins 1 to 4, respectively.

\begin{table}[!htb]
	\caption{Efficiency-corrected signal yields $Y^{K3\pi}_i$ and $Y^{K\pi\pi^0}_i$ in bin $i$ of $D \to \kspipi$ phase space, where the uncertainties are statistical only. The efficiencies integrated over all bins are $(14.22\pm0.04)\%$ and $(12.9\pm0.04)\%$ for $D \to \kmthreepi$ and $D \to \kmpipio$, respectively.}
	\label{tab:kspipiglobal}
	\centering
	\begin{tabular}{cccccc}\toprule
		Bin    &$\kmthreepi$ &$\kmpipio$\hspace*{0.0cm}  & \;\,Bin     &$\kmthreepi$ &$\kmpipio$\hspace*{0.0cm}\\
		\midrule
$1$                    & $17675\pm424$ & $31476\pm587$ & $-1$                   & $8536\pm281$ & $14826\pm390$ \\
$2$                    & $8311\pm283$  & $15515\pm394$ & $-2$                   & $2092\pm144$ & $3412\pm182$  \\
$3$                    & $6775\pm237$  & $11606\pm330$ & $-3$                   & $2072\pm128$ & $3731\pm180$  \\
$4$                    & $2278\pm140$  & $4095\pm199$  & $-4$                   & $1636\pm117$ & $2899\pm170$  \\
$5$                    & $7753\pm267$  & $15441\pm415$ & $-5$                   & $4697\pm207$ & $8039\pm288$  \\
$6$                    & $5478\pm237$  & $10887\pm342$ & $-6$                   & $1331\pm128$ & $2375\pm172$  \\
$7$                    & $12419\pm350$ & $21811\pm511$ & $-7$                   & $1503\pm126$ & $2090\pm164$  \\
$8$                    & $13448\pm374$ & $24236\pm534$ & $-8$                   & $2829\pm168$ & $5344\pm247$  \\
		\bottomrule
	\end{tabular}
\end{table}

\begin{table}[!htb]
	\caption{Efficiency-corrected signal yields ${Y^\prime}^{K3\pi}_i$ and ${Y^\prime}^{K\pi\pi^0}_i$ in bin $i$ of $D \to \klpipi$ phase space, where the uncertainties are statistical only. The efficiencies integrated over all bins are $(20.72\pm0.03)\%$ and $(20.55\pm0.05)\%$ for $D \to \kmthreepi$ and $D \to \kmpipio$, respectively.}
	\label{tab:klpipiglobal}
	\centering
	\begin{tabular}{cccccc}\toprule
		Bin    &$\kmthreepi$ &$\kmpipio$\hspace*{0.0cm}  & \;\,Bin     &$\kmthreepi$ &$\kmpipio$\hspace*{0.0cm}\\
		\midrule
$1$                    & $25820\pm490$ & $45006\pm668$ & $-1$                   & $13068\pm338$ & $23046\pm453$ \\
$2$                    & $11843\pm334$ & $21081\pm463$ & $-2$                   & $2854\pm179$  & $5849\pm253$  \\
$3$                    & $10032\pm291$ & $17441\pm404$ & $-3$                   & $3503\pm229$  & $5552\pm290$  \\
$4$                    & $4180\pm196$  & $7278\pm281$  & $-4$                   & $2234\pm212$  & $3816\pm250$  \\
$5$                    & $12278\pm307$ & $22966\pm489$ & $-5$                   & $7547\pm285$  & $11709\pm354$ \\
$6$                    & $8669\pm280$  & $14656\pm392$ & $-6$                   & $1891\pm185$  & $3733\pm261$  \\
$7$                    & $18411\pm415$ & $31310\pm564$ & $-7$                   & $2322\pm187$  & $4079\pm264$  \\
$8$                    & $18197\pm454$ & $33650\pm638$ & $-8$                   & $5102\pm227$  & $8733\pm300$  \\
		\bottomrule
	\end{tabular}
\end{table}

\section{Measurement of the observables and fit to the hadronic parameters}
\label{sec:obs}

A global analysis is performed, in which the $D\to\kmthreepi$ and $D \to \kmpipio$ phase spaces are treated inclusively.  In addition, a binned analysis is carried out, where the $D\to\kmthreepi$ phase space is partitioned according to the four-bin scheme defined in Ref.~\cite{Evans:2019wza}.

The determination of the $C\!P$-tag and like-sign observables, and their interpretations in terms of hadronic parameters, relies on the external inputs, including branching fractions and auxiliary parameters. The adopted values and their sources are provided in Table~\ref{tab:inputs}.

\begin{table}[!h]
	\caption{Input parameters used in the determination of the observables and hadronic parameters, where all uncertainties quoted are total.}
	\begin{center}
		\begin{tabular}{lllc}
			\toprule
			Input parameter \hspace{15mm} & Value &Reference\\
			\midrule
			$\br{}(\Dz\to \kpthreepi)/\br{}(\Dz\to \kmthreepi)$ & (3.22$\pm$0.05)$\times 10^{-3}$ &\hspace*{0.5cm}\cite{ParticleDataGroup:2024cfk}\\
			$\br{}(\Dz\to \kppipio)/\br{}(\Dz\rightarrow \kmpipio)$ & (2.12$\pm$0.07)$\times 10^{-3}$ &\hspace*{0.5cm}\cite{ParticleDataGroup:2024cfk}\\
			$\dkpi$ &$(190.2\pm2.8)^{\circ}$ &\hspace*{0.5cm}\cite{LHCb-CONF-2024-004}\\
			$(\rkpi)^2$ &(0.344$\pm$0.002)$\times 10^{-2}$
			&\hspace*{0.5cm}\cite{LHCb-CONF-2024-004}\\
			$x$		   &($0.40\pm0.05$)$\%$
			&\hspace*{0.5cm}\cite{LHCb-CONF-2024-004}\\
			$y$		   &$(0.636^{+0.020}_{-0.019})\%$ &\hspace*{0.5cm}\cite{LHCb-CONF-2024-004}\\
			$F_+^{\pi\pi\pi^0}$ &0.9406$\pm$0.0042 &\hspace*{0.5cm}\cite{BESIII:2024nnf}\\
			\bottomrule
		\end{tabular}
	\end{center}
	\label{tab:inputs}
\end{table}

When fitting the signal yields tagged by $D\to K_{S(L)}^0\pi^+\pi^-$ in bins of phase space, ${Y^{(\prime)}}^{K3\pi}_i$ and ${Y^{(\prime)}}^{K\pi\pi^0}_i$, to the expected distribution of events, expressed in Eqs.~(\ref{eq_kspipi}) and (\ref{eq_klpipi}), respectively, it is necessary to know the strong-phase parameters $c_i^{(\prime)}$ and $s_i^{(\prime)}$, defined in Eq.~(\ref{eq:ci}), as well as the $K_i^{(\prime)}$ parameters, defined in Eq.~(\ref{eq:ki}). 
Values for these parameters are taken from the measurement performed by the BESIII Collaboration~\cite{BESIII:2025nsp}. Notably, the $K_i^{(\prime)}$ parameters are recalculated with the contributions from $D\to\kmthreepi$ and $\kmpipio$ excluded.

\subsection{Global analysis}
\subsubsection{Determination of the $C\!P$-tag and like-sign observables}

The $C\!P$-tag observables, $\rho_{C\!P\pm}^{K3\pi}$ and $\rho_{C\!P\pm}^{K\pi\pi^0}$, are determined for each tag according to Eq.~(\ref{eq:kpinorm}). These calculations use the efficiency-corrected yields of $C\!P$-tagged signal and the $D \to K^-\pi^+$ mode listed in Table~\ref{tab:cpyields}. Additionally, the calculations incorporate the single-tag yields of the decays $D \to K^-\pi^+,~D\to\kmpipio$, and $D\to\kmpipipi$, as reported in Ref.~\cite{BESIII:2023exq}. Correction factors of $\rho_{C\!P+}^{K\pi} = 1.126 \pm 0.001$ and $\rho_{C\!P-}^{K\pi} = 0.882 \pm 0.001$ are applied, derived from external parameters $x$, $y$, $r_D^{\rm K\pi}$ and $\delta_D^{K\pi}$~\cite{LHCb-CONF-2024-004}. The resulting values are presented in Tables~\ref{tab:cp_k3pi} and~\ref{tab:cp_kpipi0}, and visualized in Figure~\ref{fig:rhocp}. The results for each signal mode and tag category are found to be consistent and are therefore combined to obtain a single averaged value for each observable. Correlations among systematic uncertainties are fully taken into account in the combination procedure.  
The like-sign observables $\rho^{K3\pi}_{LS}$,  $\rho^{K3\pi}_{K\pi, LS}$, $\rho^{K3\pi}_{K\pi\pi^0, LS}$,  $\rho^{K\pi\pi^0}_{LS}$, and $\rho^{K\pi\pi^0}_{K\pi, LS}$ are determined according to Eqs.~(\ref{eq:rho_ls2}) and~(\ref{eq:rho_lsx2}), using the yields of like-sign and opposite-sign double tags listed in Table~\ref{tab:flavouryields}. The results are summarized in Table~\ref{tab:rhoresults}.
The values of $\rho^{K3\pi}_{C\!P\pm}$, $\rho^{K\pi\pi^0}_{C\!P\pm}$, and several of the like-sign observables are incompatible with unity, indicating the presence of significant quantum-correlation effects.

\begin{table}[!htb]
	\caption{Results for $\rho_{C\!P\pm}^{K3\pi}$ shown for individual tags and the combined averages, where the first and second uncertainties are statistical and systematic, respectively.}
	\begin{center}
		\begin{tabular}{lclc}
			\toprule
			$C\!P$-even tag		&$\rho^{K3\pi}_{C\!P+}$    & $C\!P$-odd tag		&$\rho^{K3\pi}_{C\!P-}$  \\
			\midrule
                $\kk$        & $1.064\pm0.022\pm0.018$    & $\kspio$                         & $0.942\pm0.019\pm0.016$    \\
$\pipi$     & $1.160\pm0.039\pm0.019$    & $\ks\eta(\gamma\gamma)$          & $0.915\pm0.053\pm0.015$    \\
$\pipipio$  & $1.066\pm0.016\pm0.018$    & $\ks\eta(\pi\pi\pi^0)$           & $0.857\pm0.091\pm0.016$   \\
$\kspiopio$ & $1.088\pm0.039\pm0.018$    & $\ks\eta^{\prime}(\eta\pi\pi)$   & $0.922\pm0.085\pm0.017$    \\
$\klpio$    & $1.021\pm0.027\pm0.017$    & $\ks\eta^{\prime}(\gamma\pi\pi)$ & $0.950\pm0.057\pm0.015$   \\
$\klomega$  & $1.097\pm0.032\pm0.017$    & $\ksomega(\pi\pi\pi^0)$          & $1.012\pm0.045\pm0.017$    \\
&              & $\ks\phi$                        & $1.057\pm0.155\pm0.016$   \\
&              & $\klpiopio$                      & $0.961\pm0.042\pm0.016$    \\
			\midrule
			Average & $1.070 \pm 0.010 \pm 0.018$ & 
			& $0.948 \pm 0.015 \pm 0.016$  \\
			\bottomrule
		\end{tabular}
	\end{center}
	\label{tab:cp_k3pi}
\end{table}

\begin{table}[!htb]
	\caption{Results for  $\rho_{C\!P\pm}^{K\pi\pi^0}$ shown for individual tags and the combined averages, where the first and second uncertainties are statistical and systematic, respectively.}
	\begin{center}
		\begin{tabular}{lclc}
			\toprule
			$C\!P$-even tag		&$\rho^{K\pi\pi^0}_{C\!P+}$    & $C\!P$-odd tag		&$\rho^{K\pi\pi^0}_{C\!P-}$  \\
			\midrule
                $\kk$        & $1.021\pm0.019\pm0.017$    & $\kspio$                         & $0.961\pm0.018\pm0.017$    \\
$\pipi$     & $1.121\pm0.034\pm0.018$    & $\ks\eta(\gamma\gamma)$          & $0.913\pm0.047\pm0.016$    \\
$\pipipio$  & $1.082\pm0.015\pm0.018$    & $\ks\eta(\pi\pi\pi^0)$           & $0.840\pm0.083\pm0.015$   \\
$\kspiopio$ & $1.131\pm0.036\pm0.018$    & $\ks\eta^{\prime}(\eta\pi\pi)$   & $0.868\pm0.072\pm0.016$    \\
$\klpio$    & $1.037\pm0.022\pm0.018$    & $\ks\eta^{\prime}(\gamma\pi\pi)$ & $0.901\pm0.050\pm0.015$    \\
$\klomega$  & $1.033\pm0.031\pm0.018$    & $\ksomega(\pi\pi\pi^0)$          & $1.001\pm0.040\pm0.017$    \\
&              & $\ks\phi$                        & $1.045\pm0.138\pm0.016$   \\
&              & $\klpiopio$                      & $0.998\pm0.034\pm0.017$    \\
			\midrule
Average     & $1.063\pm0.009\pm0.018$    &                                  & $0.958\pm0.013\pm0.016$    \\
			\bottomrule
		\end{tabular}
	\end{center}
	\label{tab:cp_kpipi0}
\end{table}

\begin{figure}[!ht]
	\begin{center}
		\includegraphics[width=.48\textwidth]{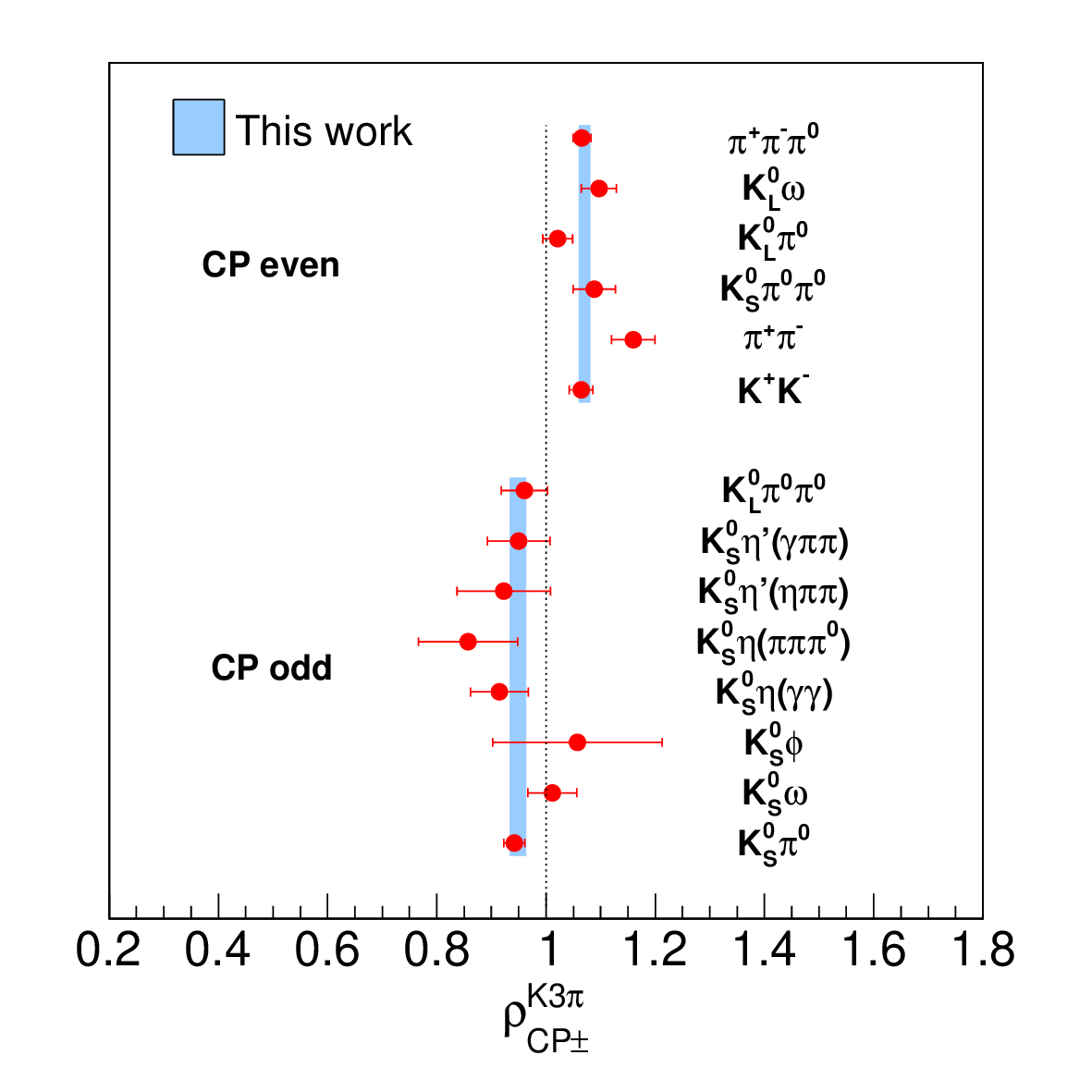}
		\includegraphics[width=.48\textwidth]{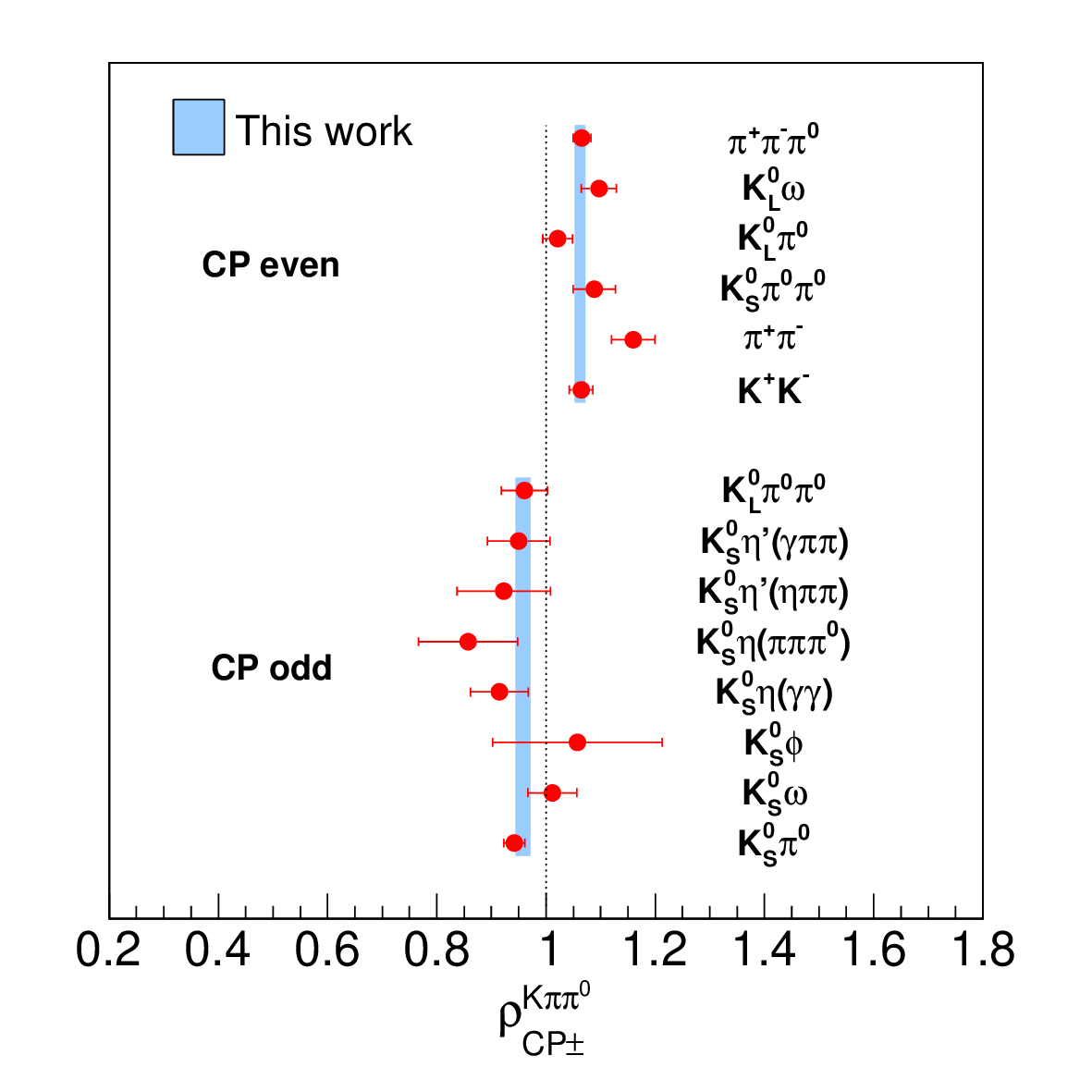}
	\end{center}
	{\caption{The observables $\rho^S_{C\!P+}$ and $\rho^S_{C\!P-}$ for (left) $S=\kmthreepi$ and (right) $S=\kmpipio$. The error bars represent the combined statistical and systematic uncertainties added in quadrature. The blue bands indicate the $\pm$1$\sigma$ region of the averaged value for each observable.}\label{fig:rhocp}}
\end{figure}

\begin{table}[!ht]
  \caption{Measured like-sign global observables for the modes $D \to \kmthreepi$ and $D \to \kmpipio$, where the first and second uncertainties are statistical and systematic, respectively.}\label{tab:rhoresults}
	\begin{center}
		\begin{tabular}{lc c lc}
			\toprule
			Observable  &  Value & & Observable & Value \\ \midrule
			$\rho^{K3\pi}_{LS}$           & $0.834 \pm 0.104 \pm 0.034$ & & $\rho^{K\pi\pi^0}_{LS}$ & $0.375 \pm 0.049 \pm 0.030$ \\
			$\rho^{K3\pi}_{K\pi,LS}$      & $0.434 \pm 0.062 \pm 0.022$ & & $\rho^{K\pi\pi^0}_{K\pi,LS}$ & $0.253 \pm 0.032 \pm 0.016$ \\
			$\rho^{K3\pi}_{K\pi\pi^0,LS}$ & $0.743 \pm 0.060 \pm 0.027$ & & & \\ 
			\bottomrule
		\end{tabular}
	\end{center}
\end{table}

\subsubsection{Assignment of systematic uncertainties}
\label{sec:syst}

The systematic uncertainties reported in Table~\ref{tab:rhoresults} are estimated by varying each input according to a Gaussian distribution with its assigned uncertainty as the width, and evaluating the resulted shifts in the central values. These shifts are then combined, accounting of any existing correlations.

There is a single source of uncertainty for the observables $\rho_{C\!P\pm}^{K3\pi}$ and $\rho_{C\!P\pm}^{K\pi\pi^0}$ arising from the selection efficiency.  
As indicated by Eq.~(\ref{eq:kpinorm}), it is necessary to know the ratio of selection efficiencies, which are obtained from Monte Carlo simulation. Associated uncertainties are derived from dedicated comparisons between data and Monte Carlo simulation~\cite{Ablikim:2018frk}, and are assigned as $0.5\%$ per charged track for reconstruction, $0.5\%$ for pion and kaon particle identification, and $1\%$ for $\pi^0$ reconstruction.
These are considered together with the smaller contribution due to the finite size of the simulated samples.

The like-sign observables are constructed by normalising the like-sign yields to the opposite-sign yields in the same final states, while accounting for differences in selection efficiencies between the CF and DCS $D \to \kmthreepi$ decays, as well as the effects of quantum correlations. The correction for quantum correlation is re-evaluated by varying the global strong-phase difference according to the result of this analysis, and the resulting change is found to be negligible. For $D \to K^-\pi^+\pi^0$, the dominant source of systematic uncertainty in simulation arises from the use of a phase-space model for the DCS process. To estimate the associated uncertainty, the decay is simulated according to the resonance structures reported by the Belle experiment in Ref.~\cite{BaBar:2008xkf} and compared with the phase-space model. A difference of approximately 2\% is observed and is assigned as a systematic uncertainty per $D \to K^-\pi^+\pi^0$ in each side of $D^0\bar{D}^0$ pair in the like-sign mode.

The yields of fully reconstructed tag modes are determined from the fits to the $M_{\rm BC}$ distributions of each signal decay. To evaluate the associated systematic uncertainty, comparisons are made with the results obtained using a two-dimensional fit to the $M_{\rm BC}$ distributions of both the signal and tag decays, taking the $D \to \kspio$ tag mode as an example. Small differences of $1.5\%$ and $0.8\%$ are assigned as the uncertainties for $\rho^{K3\pi}_{C\!P\pm}$ and $\rho^{K\pi\pi0}_{C\!P\pm}$, respectively. No uncertainty is assigned to the like-sign observables, as any potential bias cancels between the like-sign yields and the corresponding opposite-sign yields used for normalisation.

The main background in $D\to \kmpipipi$ arises from the decay $D\to \ks K^-\pi^+$, particularly in the case of like-sign double tags. The uncertainty on this background contribution is evaluated by incorporating knowledge of the decay’s coherence factor and average strong-phase difference, measured to be $0.70 \pm 0.08$ and $(0.1 \pm 15.7)^\circ$, respectively~\cite{Insler:2012pm}. These parameters are required to correct for quantum-correlation effects in the decay rate. Additional uncertainties arising from the branching fraction~\cite{ParticleDataGroup:2024cfk} and the selection efficiency are also taken into account.
This background component represents the dominant source of systematic uncertainty in the like-sign tag analysis.

A final source of uncertainty affecting the $C\!P$-tag and like-sign observables arises from the limited precision of the input parameters, whose values are taken from the measurements listed in Table~\ref{tab:inputs}.

When analyzing the efficiency-corrected signal yields tagged by $D\to\kslpipi$ in bins of phase space, ${Y^{(\prime)}}^{K3\pi}_i$ and ${Y^{(\prime)}}^{K\pi\pi^0}_i$, it is sufficient to control the relative efficiency variation. This variation is less than 15\% and taken from Monte Carlo simulation. Any potential bias in these efficiency corrections is assumed to be negligible. Uncertainties accounting for the finite size of the simulation samples are still considered. The fits to the expected event distributions incorporate uncertainties in the strong-phase parameters and in the $K_i^{(\prime)}$ values for $D \to \kslpipi$, as reported in Ref.~\cite{BESIII:2025nsp}.

\subsubsection{Fit to the hadronic parameters}
\label{sec:globalhadronic}

A $\chi^2$ fit is performed to the full set of $C\!P$-tagged, like-sign, and $D\to\kslpipi$-tagged observables defined in Eqs.~(\ref{eq:kpinorm}), (\ref{eq:rho_ls2}), (\ref{eq:rho_lsx2}) and \ref{eq_kspipi}, \ref{eq_klpipi} in order to extract the underlying physics parameters: $R_{K3\pi}$, $\delta_D^{K3\pi}$, $r_D^{K3\pi}$, $R_{K\pi\pi^0}$, $\delta_D^{K\pi\pi^0}$ and $r_D^{K\pi\pi^0}$.
Each individual $C\!P$-tag observable is treated as an independent measurement in the fit, resulting in a total of 97 observables and 6 free fit parameters. 
Systematic uncertainties associated with external inputs are not included directly in the uncertainties of the observables. 
Instead, auxiliary parameters $\delta_D^{K\pi}$, $r_D^{K\pi}$, $x$, $y$, $F_+^{\pi\pi\pi^0}$, $K^{(\prime)}_i$, $c^{(\prime)}_i$, and $s^{(\prime)}_i$ are treated as nuisance fit parameters and constrained with Gaussian penalty terms in the $\chi^2$ function, using their measured central values and covariance matrices. When terms involve the ratio of branching fractions, such as $\br{} (D^0 \to  K^+\pi^-)/\br{} (D^0 \to K^-\pi^+)$,
$\br{} (D^0 \to \kpthreepi) / \br{} (D^0 \to \kmthreepi)$ or $\br{} (D^0 \to \kppipio) / \br{} (D^0 \to \kmpipio)$, these ratios are replaced by the corresponding theoretical expressions (see Eqs.~(\ref{eq:BRCF}) and~(\ref{eq:BRDCS})). Additional terms are included in the $\chi^2$ to constrain the measured values of 
$\br{} (D^0 \to \kpthreepi) / \br{} (D^0 \to \kmthreepi)$ and $\br{} (D^0 \to \kppipio) / \br{} (D^0 \to \kmpipio)$ to their theoretical predictions. In summary, the function minimised in the fit is
\begin{equation}
\chi^2 = \chi^2_{C\!P} + \chi^2_{LS} + \chi^2_{K^0_{S,L} \pi\pi} + \chi^2_{\rm aux}\, ,
\label{eq:chi2fn}
\end{equation}
where $\chi^2_{C\!P}$, $\chi^2_{LS}$ and $\chi^2_{K^0_{S,L} \pi\pi}$ denote the contributions from the $C\!P$ tags, like-sign tags, and $\kslpipi$ tags, respectively. The term $\chi^2_{\rm aux}$ accounts for the Gaussian constrains on auxiliary parameters and branching fraction ratios.
Validation studies using simulated datasets confirm that the fit is unbiased.

The fit converges with a $\chi^2/{\rm n.d.f.}=116.1/105$, indicating good overall agreement among the inputs. Figure~\ref{fig:kspipi_results} presents the measured observables ${Y^{(\prime)}}^{K3\pi}_i$ and ${Y^{(\prime)}}^{K\pi\pi^0}_i$, overlaid with the fit predictions. For comparison, predictions under the uncorrelated hypothesis, which are directly proportional to the $K^{(\prime)}_i$ values, are also shown.

\begin{figure*}[!htb]
	\centering
    \subfigure{
	\includegraphics[width=.48\textwidth]{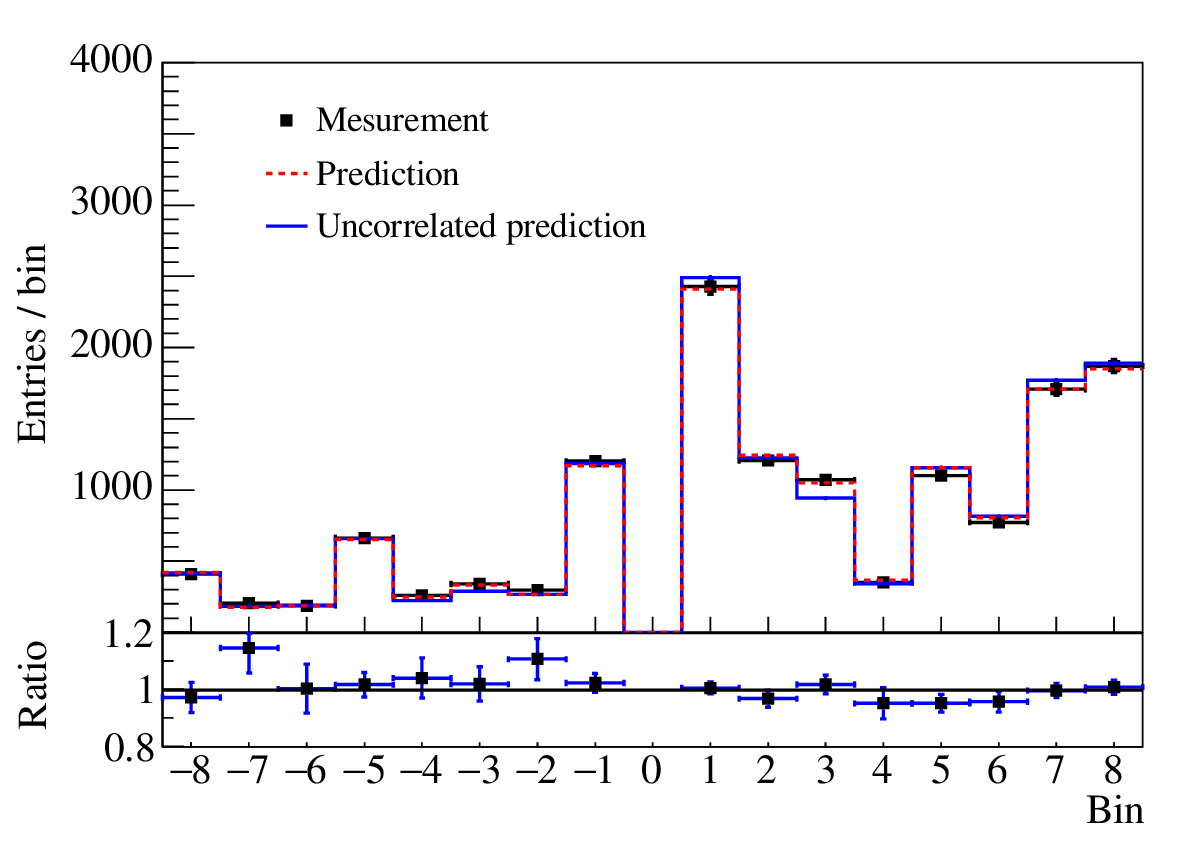}
    \put(-50,120){(I)}
    }
    \subfigure{
	\includegraphics[width=.48\textwidth]{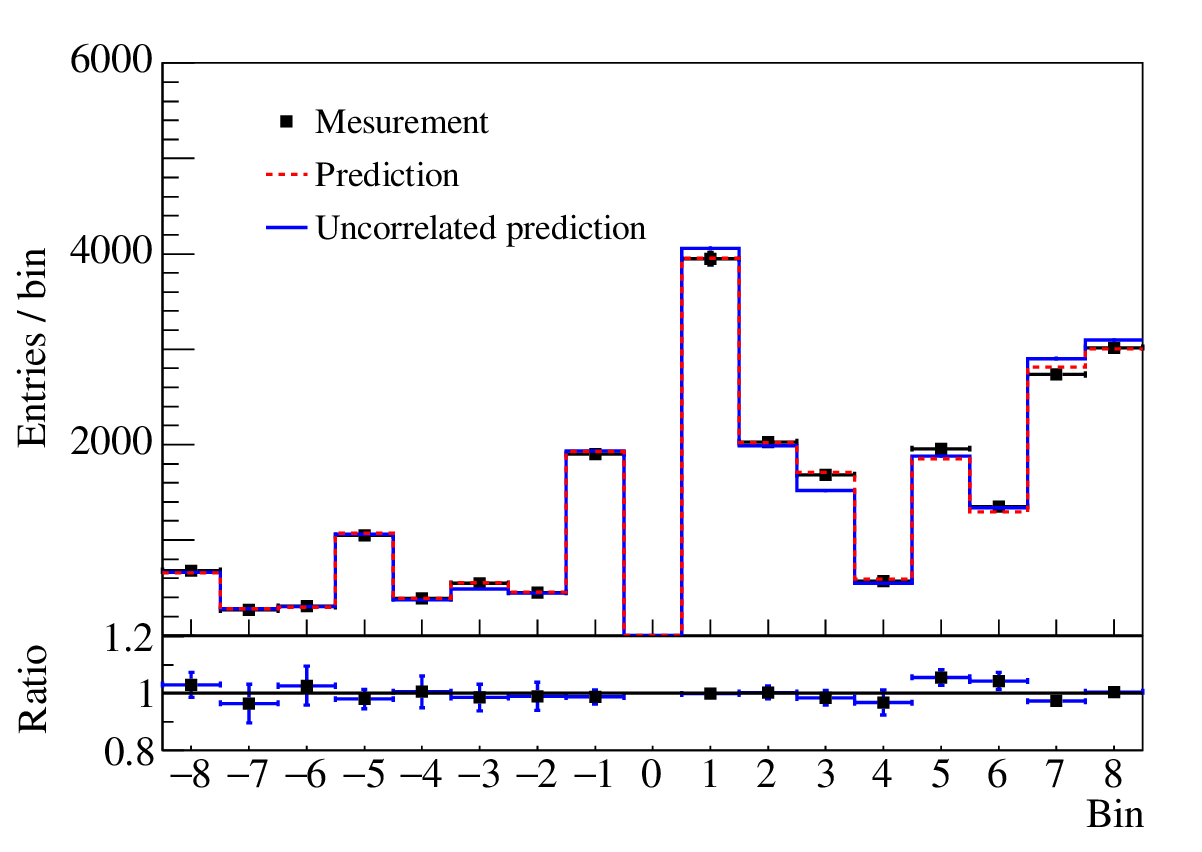}
    \put(-50,120){(II)}}
    \subfigure{
	\includegraphics[width=.48\textwidth]{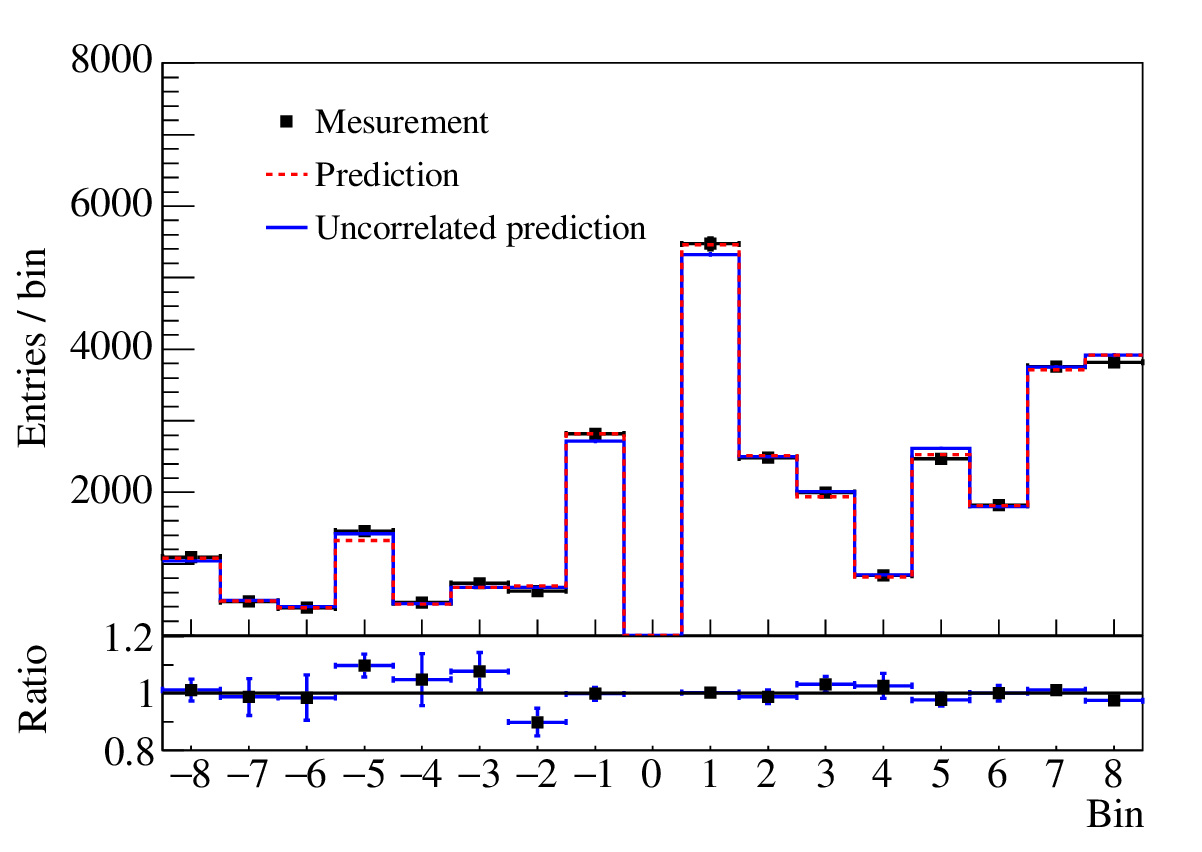}
    \put(-50,120){(III)}
    }
    \subfigure{
	\includegraphics[width=.48\textwidth]{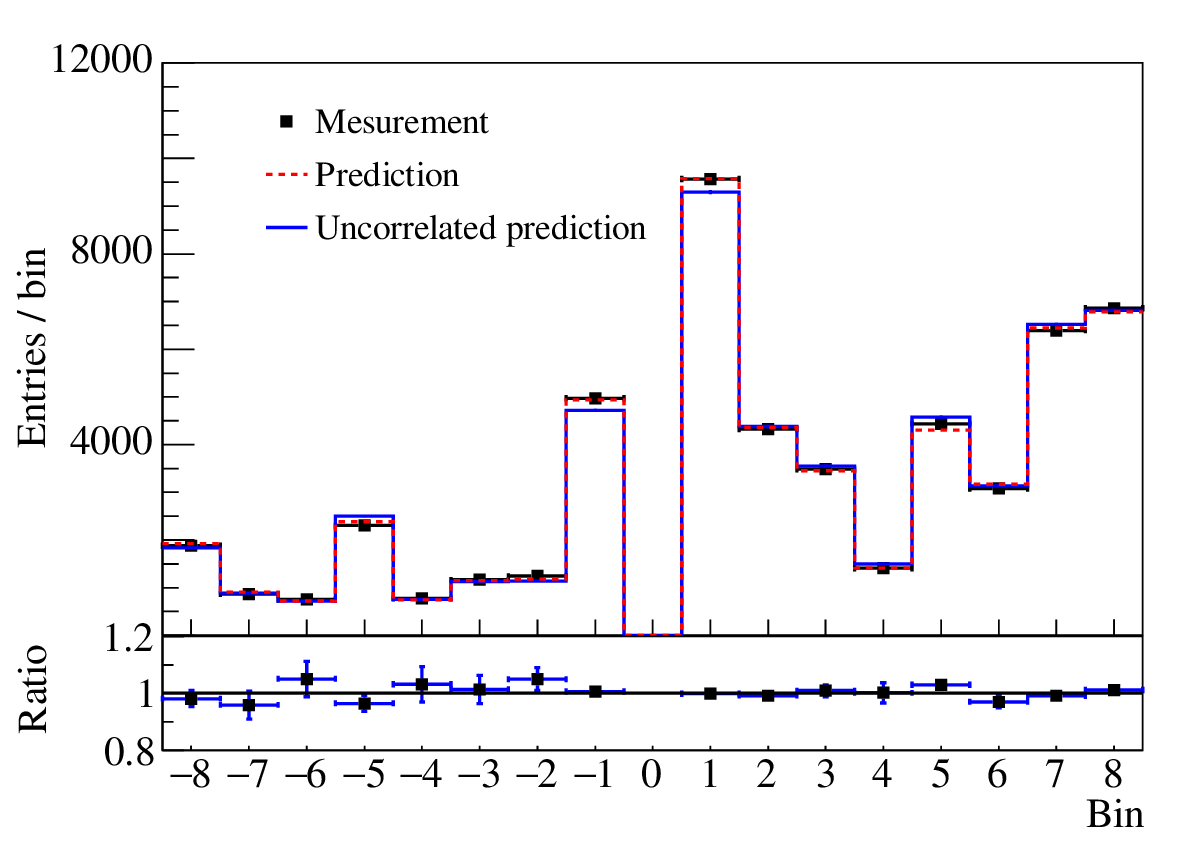}
    \put(-50,120){(IV)}
    }
	\caption{The top panels show (I) $Y^{K3\pi}_i$,  (II) $Y^{K\pi\pi^0}_i$, and (III) ${Y^{\prime}}^{K3\pi}_i$, (IV) ${Y^{\prime}}^{K\pi\pi^0}_i$ observables for the $D\to\kslpipi$ tags, with the predictions from the fit to all the BESIII inputs. Also shown are the expectations from the uncorrelated hypothesis. The bottom panels show the ratio of the measurement to the prediction in each bin.}
	\label{fig:kspipi_results}
\end{figure*}

The fitted values of the hadronic parameters are listed in Table~\ref{tab:parameter}, while the $\Delta \chi^2$ contours in the $(\Rkthreepi,\dkthreepi)$ and $(\Rkpipio,\dkpipio)$ parameter space are shown in Figure~\ref{fig:parameter1_1sigma}. 
 The measured value of $\Rkthreepi$ is consistent with the prediction of $0.46$ from the LHCb amplitude models~\cite{Aaij:2017kbo}. Compared to the previous measurement~\cite{BESIII:2021eud}, the present results provide significantly improved constraints. 
 The fit returns values for all auxiliary parameters that are consistent with their experimentally measured inputs.

\begin{table}[!ht]
	\caption{Fitted central values for the hadronic parameters, from the global and binned analyses.   The uncertainties include both statistical and systematic contributions.}
	\label{tab:parameter}
	\begin{center}
		\begin{tabular}{lccccccc}
			\toprule
			Parameter       & \hspace*{0.3cm}Global fit\hspace*{0.3cm}   & \multicolumn{4}{c}{Binned fit} \\
			&   &Bin 1 &Bin 2 & Bin 3 &Bin 4 \\
			\midrule
$R_{K3\pi}$                      & $0.51\pm0.04$     & $0.72_{-0.14}^{+0.14}$ & $0.65_{-0.07}^{+0.08}$ & $0.80_{-0.06}^{+0.07}$ & $0.54_{-0.19}^{+0.18}$ \\ \\
$\delta_D^{K3\pi}(^\circ)$       & $182^{+14}_{-13}$ & $124_{-8}^{+10}$       & $170_{-12}^{+14}$      & $179_{-11}^{+11}$      & $260_{-13}^{+10}$      \\ \\
$r_D^{K3\pi}(\times10^{-2})$     & $5.51\pm0.05$     & $5.48\pm0.05$          & $5.96\pm0.05$          & $5.86\pm0.05$          & $4.83\pm0.04$          \\ \\
$R_{K\pi\pi^0}$                  & $0.75\pm0.04$     & \multicolumn{4}{c}{$0.75\pm0.03$}                                                                 \\ \\
$\delta_D^{K\pi\pi^0}(^\circ)$   & $206_{-14}^{+9}$  & \multicolumn{4}{c}{$209_{-8}^{+7}$}                                                               \\ \\
$r_D^{K\pi\pi^0}(\times10^{-2})$ & $4.40\pm0.09$     & \multicolumn{4}{c}{$4.41\pm0.08$}                                                                 \\
			\bottomrule
		\end{tabular}
	\end{center}
\end{table}

\begin{figure}[!ht]
	\begin{center}
		\includegraphics[width=.48\textwidth]{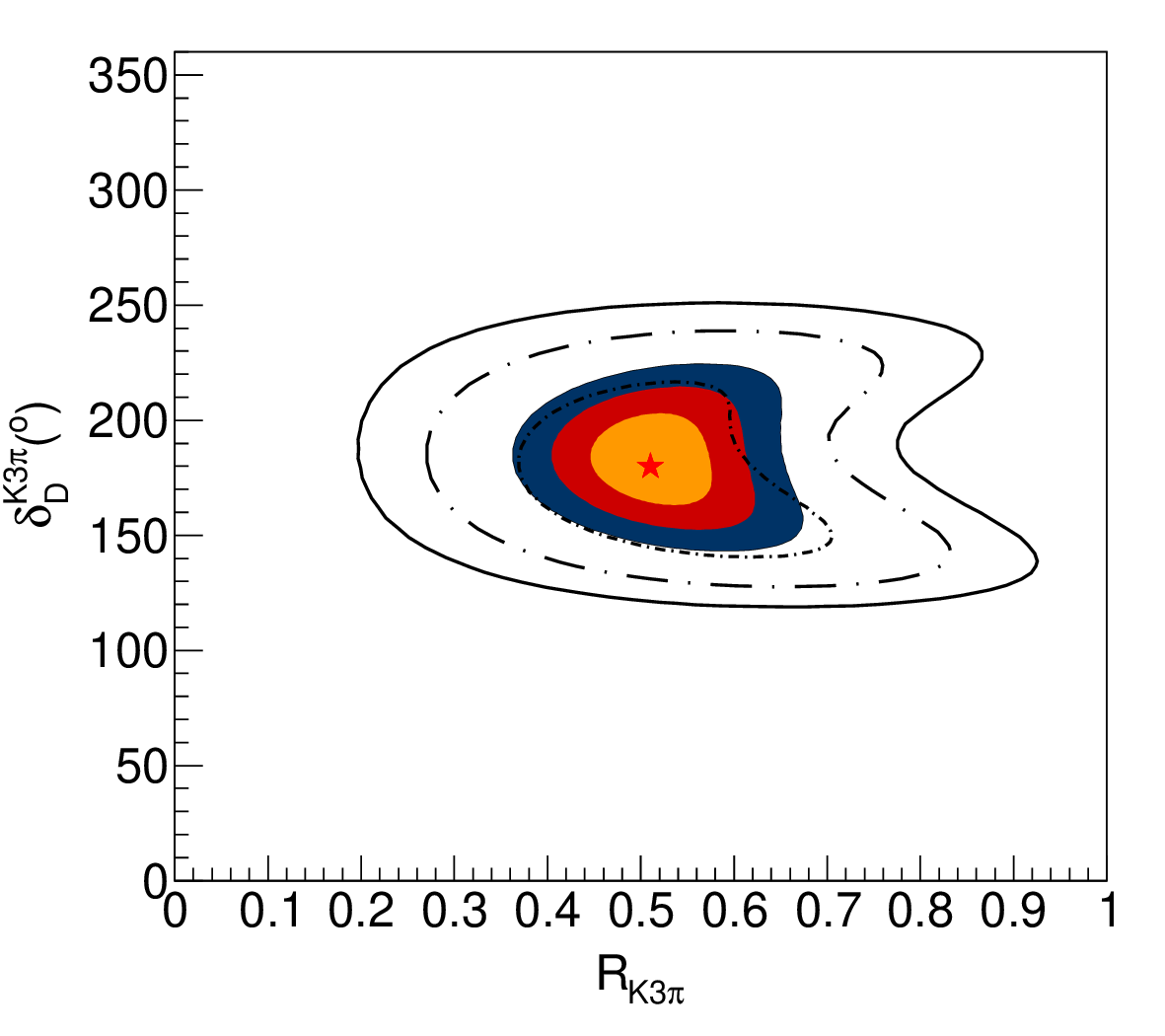}
		\includegraphics[width=.48\textwidth]{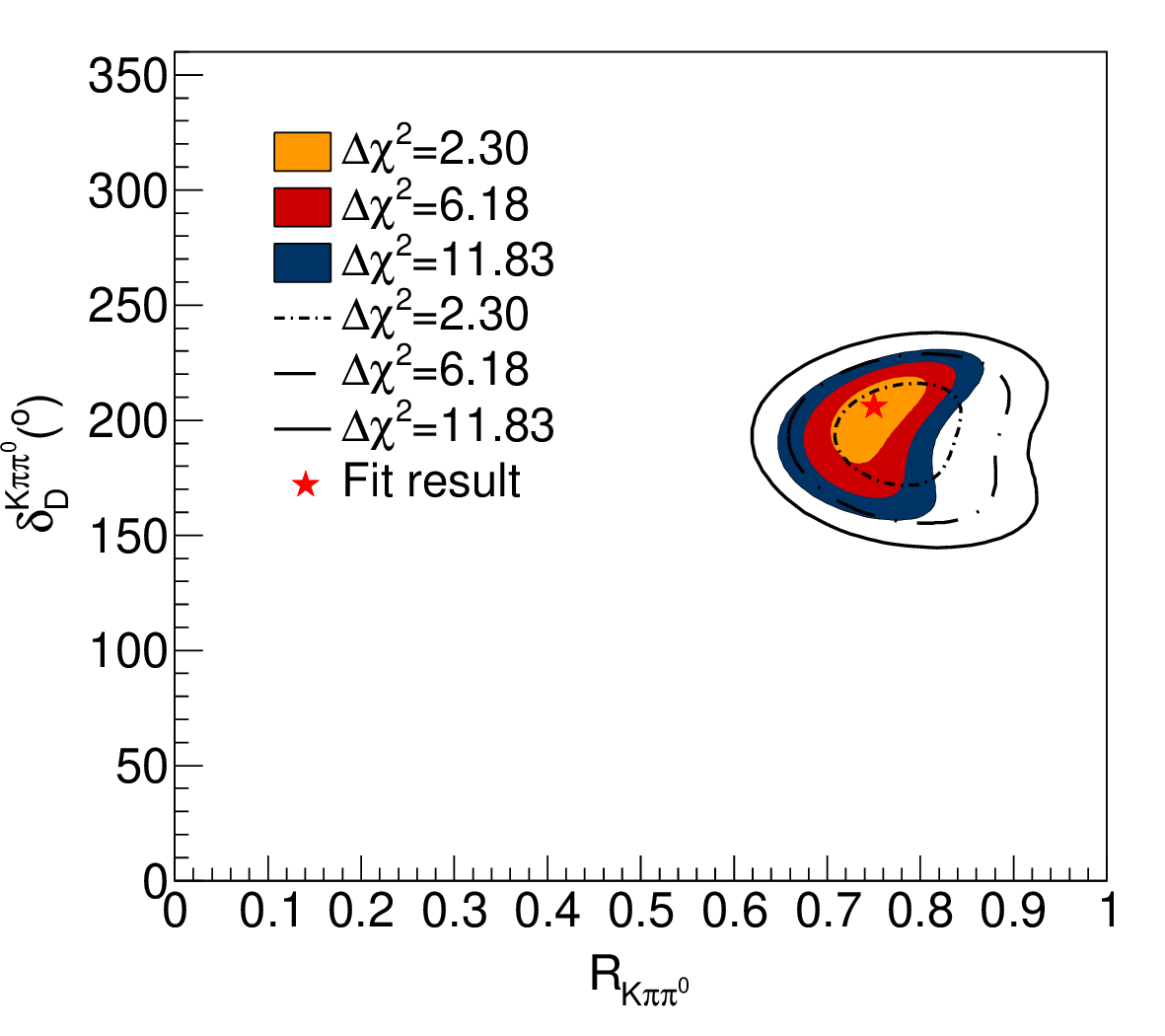}
		\caption{Scans of $\Delta \chi^2$ in the global ($\Rkthreepi$, $\dkthreepi$) and ($\Rkpipio$, $\dkpipio$) parameter space, where the filled color contours represent the current results and the various black lines indicate the previous results from the BESIII Collaboration~\cite{BESIII:2021eud}. The $\Delta \chi^2$ = 2.30, 6.18, and 11.83 contours correspond to the 68.3\%, 95.4\%, and 99.7\% confidence levels in the two-dimensional parameter space, respectively.}
		\label{fig:parameter1_1sigma}
	\end{center}
\end{figure}

To assess the intrinsic statistical precision of the sample and quantify the impact of each systematic input on the overall uncertainty, a series of fits is performed. In each iteration, one input is varied according to a Gaussian distribution with a width set by its assigned uncertainty, while all other inputs are fixed. The standard deviation of the resulting distribution of the fitted parameters is taken as the uncertainty associated with that particular input. The results of this study are summarized in Table~\ref{tab:inputsys}. The most significant sources of systematic uncertainty arise from the limited size of the $C\!P$-tagged $D \to K^-\pi^+$ samples, the uncertainty on the $D \to \ks K^-\pi^+$ background, and the precision of the strong-phase parameters and the $K^{(\prime)}_i$ values for $D\to\kslpipi$. The statistical uncertainty remains the dominant contribution for all four measured parameters.

\begin{table}[!htbp]
	\caption{The estimated systematic uncertainties on the global hadronic parameters. The upper section lists contributions associated with the detector and measurement procedure, while those in the lower section arise from the external parameters. The statistical uncertainty is provided for the comparison.}
	\label{tab:inputsys}
	\begin{center}
		\begin{tabular}{lrrrr}
			\toprule
			Systematics & $\Rkthreepi$ &$\dkthreepi$ & \hspace*{0.4cm}$\Rkpipio$\hspace*{0.0cm} &$\dkpipio$\\
			\midrule
$K^\pm, \pi^\pm$ tracking and identification&$<0.01$&$0.3$&$<0.01$&$<0.1$\\
$\pi^0$ reconstruction&$<0.01$&$<0.1$&$<0.01$&$<0.1$\\
Monte Carlo statistics&$<0.01$&$0.6$&$<0.01$&$0.6$\\
$D\to\ks K\pi$ background&$<0.01$&$0.7$&$<0.01$&$<0.1$\\
Fit method for signal yield&$0.01$&$1.0$&$<0.01$&$2.1$\\	
DCS $D\to \kmpipio$ modeling&$<0.01$&$<0.1$&$<0.01$&$0.9$\\
			\midrule
$c_i^{(\prime)},s_i^{(\prime)}$&$<0.01$&$_{-1.9}^{+1.7}$&$<0.01$&$_{-1.7}^{+1.3}$\\
$K_i^{(\prime)}$&$_{-0.02}^{+0.01}$&$_{-5.8}^{+8.8}$&$_{-0.02}^{+0.02}$&$_{-9.1}^{+3.5}$\\
Input branching fractions &$<0.01$&$_{-0.2}^{+0.2}$&$_{-0.01}^{+0.01}$&$_{-1.2}^{+1.2}$\\
$r_D^{K\pi}$&$<0.01$&$<0.1$&$<0.01$&$_{-0.2}^{+0.2}$\\
$\delta_D^{K\pi}$&$<0.01$&$_{-0.3}^{+0.7}$&$<0.01$&$_{-1.6}^{+1.4}$\\
$x,y$&$<0.01$&$_{-0.1}^{+0.1}$&$<0.01$&$_{-0.1}^{+0.1}$\\
$F^{\pi\pi\pi^0}$&$<0.01$&$<0.1$&$<0.01$&$<0.1$\\
Single tag yields&$<0.01$&$<0.1$&$<0.01$&$<0.1$\\
			\midrule
			Statistical uncertainty & $_{-0.04}^{+0.04}$&$_{-11}^{+12}$&$_{-0.03}^{+0.03}$&$_{-11}^{+8}$\\
			\bottomrule
		\end{tabular}
	\end{center}
\end{table}

\subsection{Binned ${ D \to \kmthreepi}$ analysis}

The binned analysis follows the same methodology as the global fit, but incorporates 170 observables and 15 free parameters.
Since the binning scheme is designed to exclude contributions from the 
$D \to \ks K^-\pi^+$ background, no uncertainty arises from this source.
Monte Carlo studies indicate that approximately  90\% of decays are correctly assigned to their respective bins. To account for bin migration, a correction is applied using a migration matrix derived from Monte Carlo simulation. Validation with ensembles of simulated experiments confirms that the fit procedure is unbiased and provides reliable uncertainty estimates.
The measured values of the binned observables are listed in Table~\ref{tab:rhoresults_binned}.

\begin{table}[!ht]
	\caption{Measured binned observables for $D \to \kmthreepi$, where the first uncertainties are statistical and the second systematic.}\label{tab:rhoresults_binned}
	\begin{center}
		\scriptsize
		\begin{tabular}{lrc c cc}
			\toprule
			\multicolumn{2}{l}{Observable}  &  Bin 1 &  Bin 2 & Bin 3 & Bin 4 \\ \midrule
\multicolumn{2}{l}{$\rho^{K3\pi}_{C\!P+}$} & $1.051\pm0.016\pm0.018$ & $1.090\pm0.017\pm0.018$ & $1.086\pm0.017\pm0.018$ & $1.011\pm0.015\pm0.017$\\
\multicolumn{2}{l}{$\rho^{K3\pi}_{C\!P-}$} & $0.917\pm0.021\pm0.015$ & $0.936\pm0.022\pm0.016$ & $0.863\pm0.021\pm0.015$ & $0.977\pm0.021\pm0.016$\\
			\multicolumn{2}{l}{$\rho^{K3\pi}_{K\pi,LS}$}      & $0.444\pm0.109\pm0.022$ & $0.211\pm0.079\pm0.011$ & $0.058\pm0.061\pm0.003$ & $0.486\pm0.115\pm0.025$ \\
			\multicolumn{2}{l}{$\rho^{K3\pi}_{K\pi\pi^0,LS}$} & $0.835\pm0.124\pm0.030$ & $0.492\pm0.092\pm0.018$ & $0.508\pm0.101\pm0.018$ & $0.729\pm0.118\pm0.027$ \\
			\multirow{4}{*}{$\rho^{K3\pi}_{LS}$}    & Bin 1   & $0.845\pm0.390\pm0.034$ & $0.477\pm0.205\pm0.019$ & $0.630\pm0.258\pm0.026$ & $1.550\pm0.399\pm0.063$ \\
			& Bin 2   &                         & $0.789\pm0.384\pm0.032$ & $0.653\pm0.255\pm0.027$ & $0.641\pm0.222\pm0.026$ \\
			& Bin 3   &                         &                         & $0.653\pm0.404\pm0.027$ & $0.433\pm0.208\pm0.018$ \\
			& Bin 4   &                         &                         &                         & $0.835\pm0.433\pm0.034$ \\
			\bottomrule
		\end{tabular}
	\end{center}
\end{table}

The fitted values of the hadronic parameters are reported in Table~\ref{tab:parameter}, and the $\Delta \chi^2$ contours in the $(\Rkthreepi,\dkthreepi)$ parameter space are shown in Figure~\ref{fig:binparameter}. The fit exhibits a satisfactory quality with $\chi^2/{\rm n.d.f.} = 180/155$. The fitted $K^{(\prime)}_i$ values and their correlation matrix are provided in Appendix~\ref{sec:ki}.

Amplitude models can be used to compute predictions for the coherence factor in each bin and the variation in strong phase between bins~\cite{Evans:2019wza}. By incorporating the measured value and the total uncertainty of $\dkthreepi$ from the global analysis, it is possible to derive effective predictions for the average strong-phase difference in each bin, along with their correlated uncertainties. Following this procedure, the predicted values of the coherence factors and strong-phase differences are found to be $\left(0.67,(95^{+14}_{-13})^\circ\right)$, $\left(0.85,(147^{+14}_{-13})^\circ\right)$, $\left(0.82,(188^{+14}_{-13})^\circ\right)$, and $\left(0.63,(247^{+14}_{-13})^\circ\right)$ for bins 1 to 4, respectively. These predictions are consistent with the measured values within two standard deviations.

\begin{figure}[!ht]
	\centering
	\includegraphics[width=.48\textwidth]{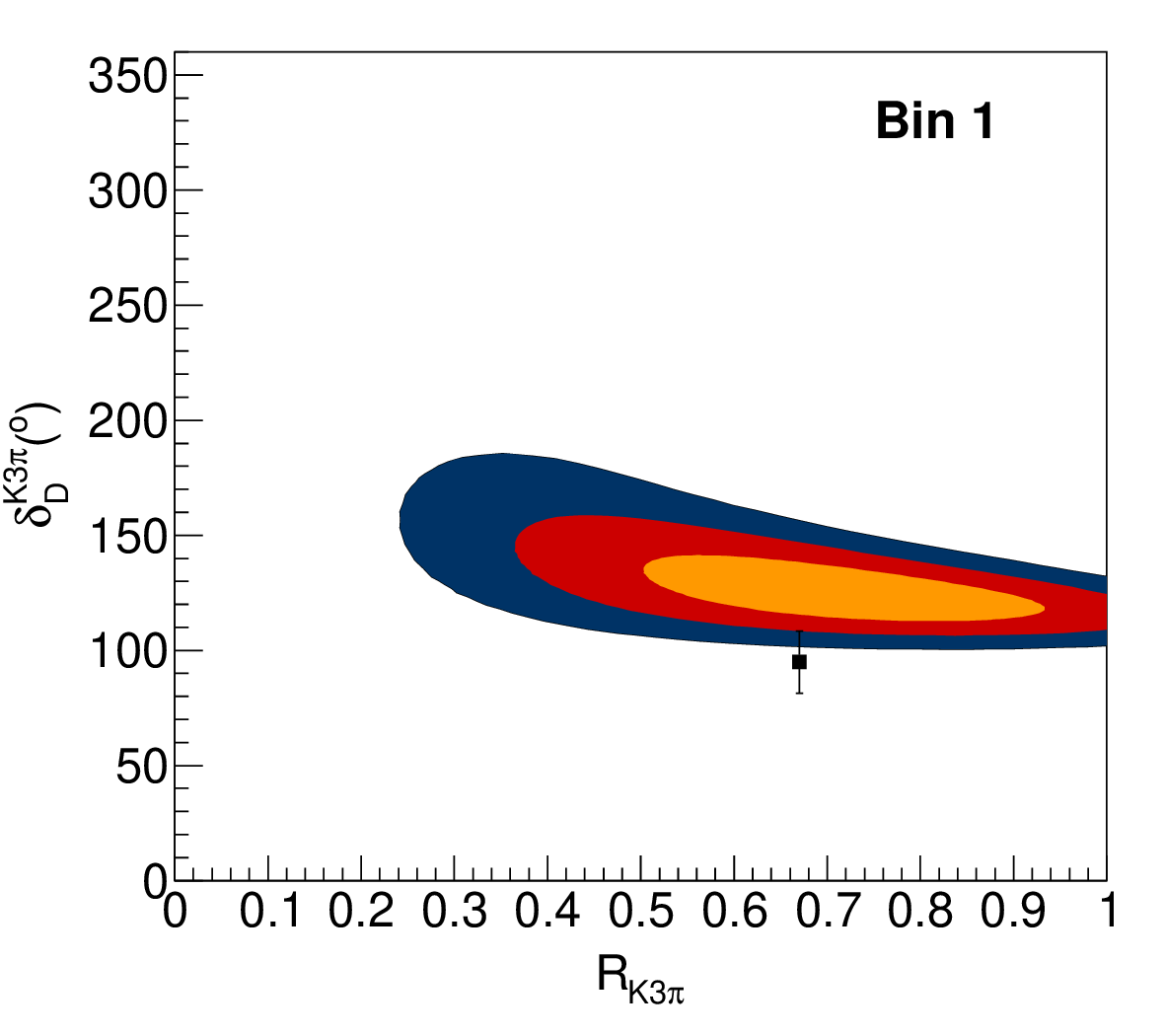}
	\includegraphics[width=.48\textwidth]{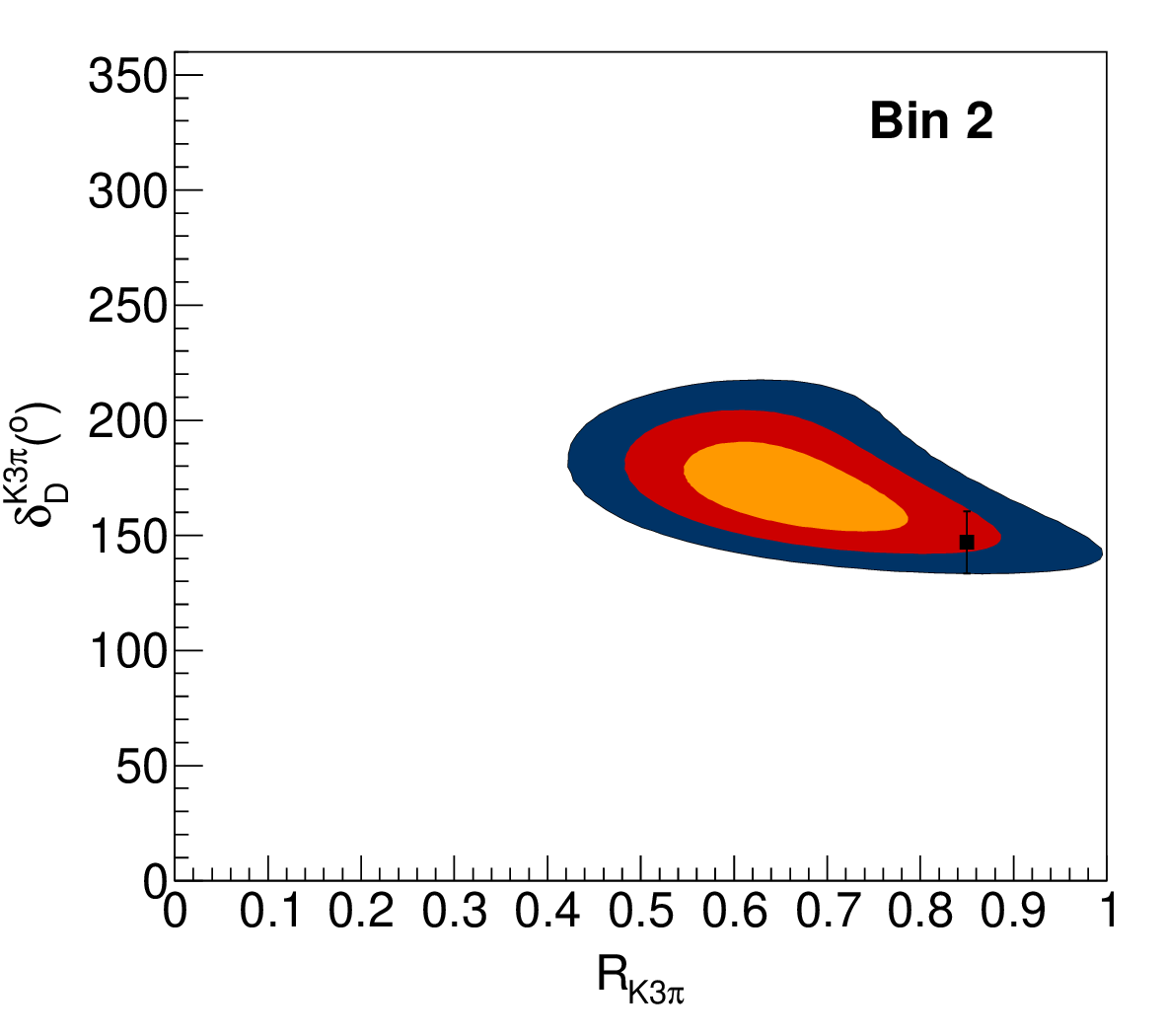}\\
	\includegraphics[width=.48\textwidth]{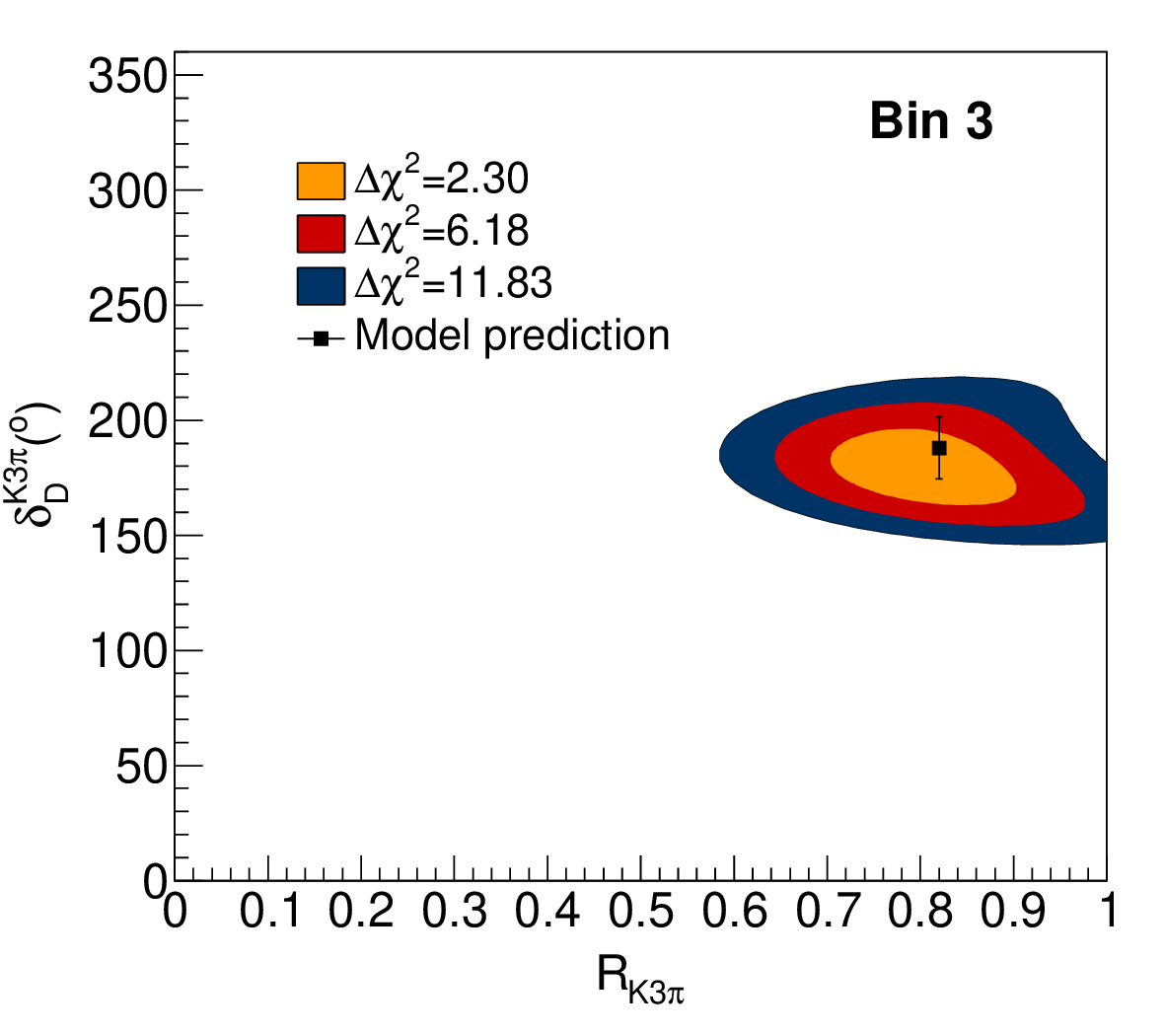}
	\includegraphics[width=.48\textwidth]{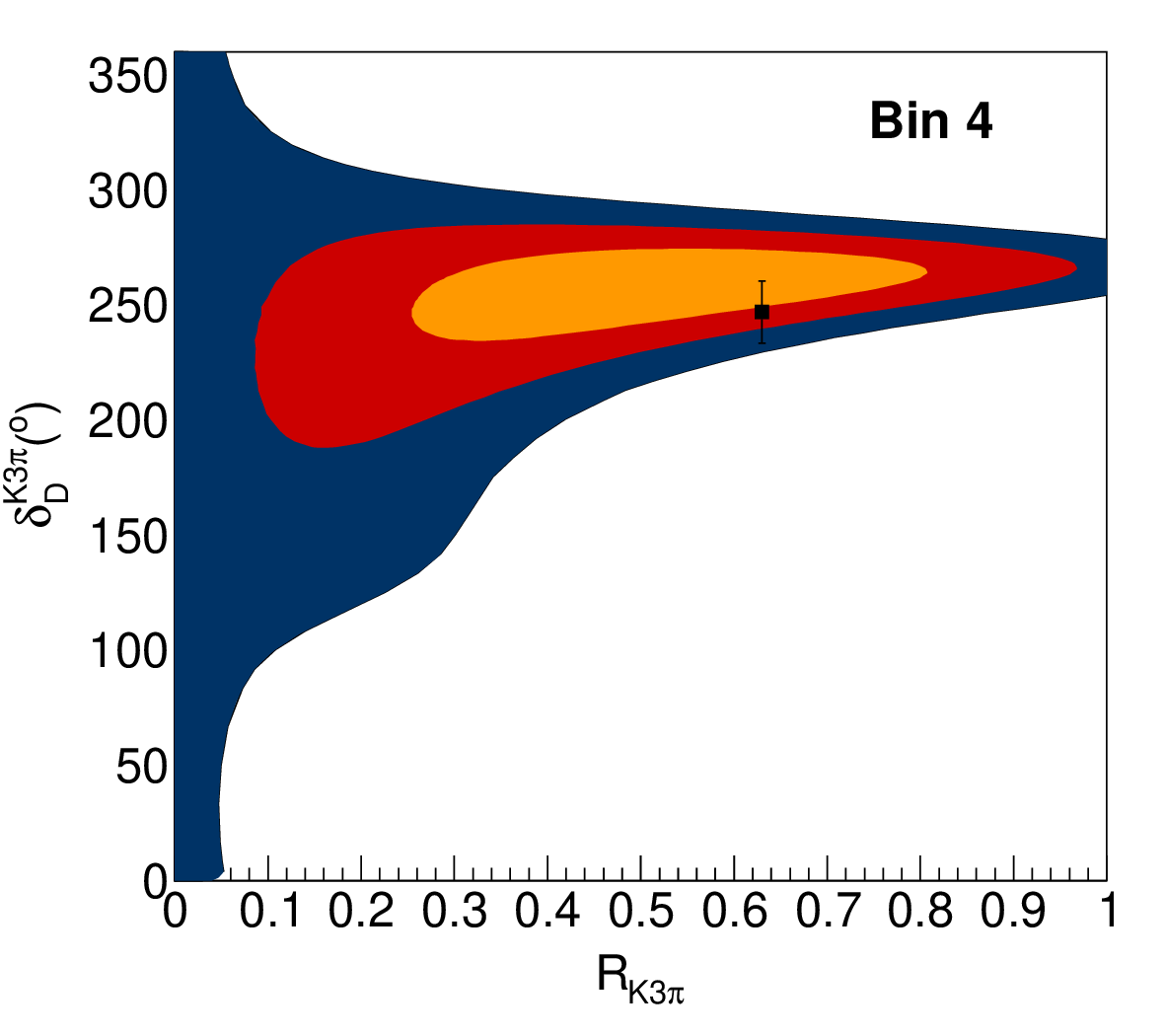}
	\caption{Scans of $\Delta \chi^2$ in the four bins of the ($\Rkthreepi$, $\dkthreepi$)  parameter space, showing the $\Delta \chi^2$=2.30, 6.18, 11.83 intervals, which correspond to the $68.3\%$, $95.4\%$ and $99.7\%$ confidence levels in the two-dimensional parameter space. The prediction from the model~\cite{LHCb:2017swu} is also indicated, where the global offset in the average strong-phase difference comes from the global result.}\label{fig:binparameter}
\end{figure}

\section{Impact of the results on the measurement of $\gamma$}
\label{sec:gamma}

Improved knowledge of the global coherence factors and average strong-phase differences for the decays $D\to \kmthreepi$ and $D \to \kmpipio$ will enhance future measurements of the Unitarity Triangle angle $\gamma$ in the $B^- \to DK^-$ decays at LHCb~\cite{Malde:2015mha,Bediaga:2018lhg} and Belle~II~\cite{Kou:2018nap}. Although neither of these channels individually provides competitive sensitivity for a stand-alone determination of $\gamma$, they offer valuable constraints when included as part of the global combination of $B^- \to DK^-$ analyses involving multiple $D$ decay modes. These constraints will be further strengthened by the BESIII measurements presented in this work.

Subdividing the phase space of $D \to \kmthreepi$ into four bins enables a stand-alone determination of $\gamma$~\cite{LHCb:2022nng}. The impact of the measurements reported in this work on such an analysis can therefore be quantified.

The values and the total uncertainties of the decay rates $\mathcal{R}_{K^{\pm}}$, measured in Ref.~\cite{LHCb:2022nng} using $9\ \rm fb^{-1}$ of Run 1 and Run 2 LHCb data, are incorporated into the binned $\chi^2$ fit, following the formalism of Ref.~\cite{LHCb:2022nng}, to extract the value of $\gamma$, the $B$-meson decay parameters, as well as the hadronic parameters of the $D$ decay. To assess the impact of the BESIII measurements on the precision of this determination, a $300\ \rm fb^{-1}$ pseudo-dataset of $B$ decay is generated based on the fitted values of $\gamma$ and the $B$-meson decay parameters. This allows the contribution of the BESIII inputs to the overall fit uncertainties to be quantified in isolation.

Figure~\ref{fig:gammastudy} shows the variation in $\Delta \chi^2$, defined as the change in $\chi^2$ relative to its minimum, as a function of $\gamma$, with the minimum centred on the input value.
An analysis of $9\ \rm fb^{-1}$ of suppressed $B$ decay data from LHCb, combined with the $D$ hadronic parameters reported in this work, is expected to determine $\gamma$ with a precision of $8.5^\circ$. The contribution from the BESIII measurements to this uncertainty is projected to be $3.5^\circ$, based on a $300\ \rm fb^{-1}$ pseudo-dataset (which has a negligible uncertainty from this large LHCb sample itself). This precision is only slightly worse than the current best stand-alone determination of $\gamma$, obtained from the LHCb analysis of $B^- \to DK^-$ with $D\to K^0_S\pi^+\pi^-$ and $D\to K^0_S K^+ K^-$, which achieves an uncertainty of approximately $5^\circ$~\cite{Aaij:2020xuj}. Therefore, the BESIII results on the binned hadronic parameters of $D \to K^-\pi^+\pi^+\pi^-$ provide an important constraint for improving the understanding of $C\!P$ violation in $b$-hadron decays.

\begin{figure}[!htb]
	\centering
	\includegraphics[width=.48\textwidth]{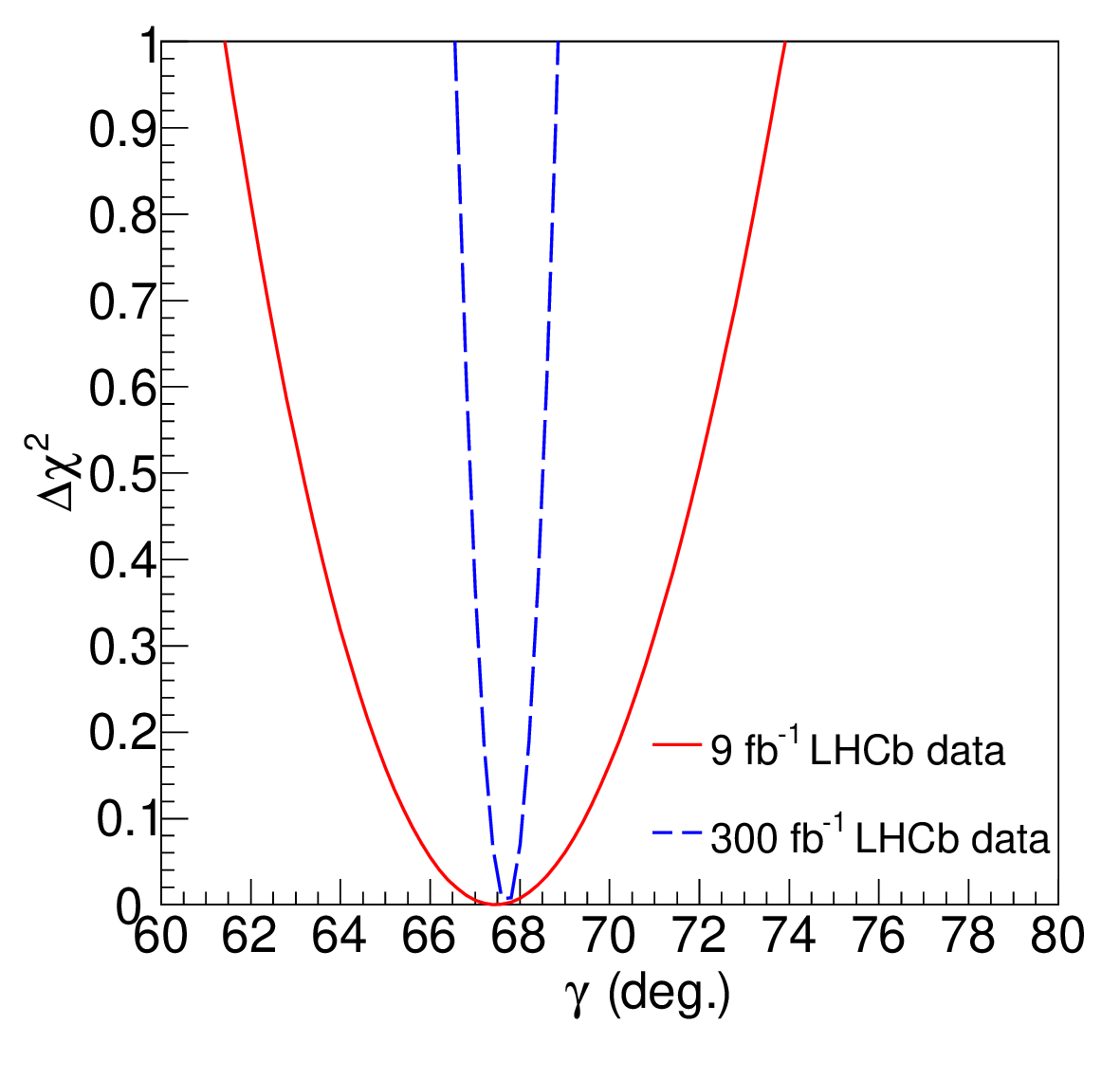}
	\caption{The distribution of $\Delta\chi^2$ vs. $\gamma$ as fitted from two ensembles of simulated data sets for $\rm 9\ fb^{-1}$ and $\rm 300\ fb^{-1}$ LHCb data.}\label{fig:gammastudy}
\end{figure}

\section{Conclusion}
\label{sec:summary}
Based on a data sample of $7.93~{\rm fb^{-1}}$ of $\ee$ annihilation collected at $\sqrt{s}=3.773\gev$ with the BESIII detector at the BEPCII collider, an analysis is performed to measure the average strong-phase difference and associated hadronic parameters of the decays $D\to\kmpipipi$ and $D\to\kmpipio$. The coherence factors and average strong-phase differences are determined to be
\begin{equation*}
\begin{aligned}
R_{K3\pi}=0.51\pm0.04,&&R_{K\pi\pi^0}=0.75\pm0.03,\\
\delta_D^{K3\pi}=\left(182^{+14}_{-13}\right)^\circ,&&\delta_D^{K\pi\pi^0}=\left(209^{+7}_{-8}\right)^\circ.
\end{aligned}
\end{equation*}
The uncertainty on the coherence factor is improved compared to the previous CLEO analysis~\cite{Evans:2016tlp} and the earlier BESIII result~\cite{BESIII:2021eud}, with the uncertainties for the $D\to\kmpipipi$ mode reduced by a factor of 2–3. The regional hadronic parameters are also determined in four bins of $D\to \kmthreepi$ phase space.
These improved hadronic parameters provide essential inputs for model-independent measurements of the CKM angle $\gamma$ and studies of charm mixing. For instance, they can refine the precision of charm mixing and $CP$ violation parameters in $D^0\to K^\pm\pi^\mp\pi^\pm\pi^\mp$ decays, as recently investigated by LHCb~\cite{LHCb:2025zgk}.
In context of the $\gamma$  measurement using $B^-\to DK^-$ decays, the BESIII measurements are expected to contribute an uncertainty of approximately $3.5^\circ$, which is smaller than the uncertainty associated with the current size of available $B$-meson decay samples at LHCb~\cite{Malde:2015mha,Bediaga:2018lhg} and Belle~II~\cite{Kou:2018nap}.
An analysis using the full BESIII dataset of $20~{\rm fb^{-1}}$ is foreseen, which will further suppress the strong-phase uncertainty.

\acknowledgments

The BESIII Collaboration thanks the staff of BEPCII (https://cstr.cn/31109.02.BEPC) and the IHEP computing center for their strong support. This work is supported in part by National Key R\&D Program of China under Contracts Nos. 2023YFA1606000, 2023YFA1606704; National Natural Science Foundation of China (NSFC) under Contracts Nos. 11635010, 11935015, 11935016, 11935018, 12025502, 12035009, 12035013, 12061131003, 12192260, 12192261, 12192262, 12192263, 12192264, 12192265, 12221005, 12225509, 12235017, 12361141819, 12405112, 12422504; the Fundamental Research Funds for the Central Universities, Lanzhou University under Contracts Nos. lzujbky-2023-stlt01, lzujbky-2025-ytB01; the Chinese Academy of Sciences (CAS) Large-Scale Scientific Facility Program; the Strategic Priority Research Program of Chinese Academy of Sciences under Contract No. XDA0480600; CAS under Contract No. YSBR-101; 100 Talents Program of CAS; The Institute of Nuclear and Particle Physics (INPAC) and Shanghai Key Laboratory for Particle Physics and Cosmology; ERC under Contract No. 758462; German Research Foundation DFG under Contract No. FOR5327; Istituto Nazionale di Fisica Nucleare, Italy; Knut and Alice Wallenberg Foundation under Contracts Nos. 2021.0174, 2021.0299; Ministry of Development of Turkey under Contract No. DPT2006K-120470; National Research Foundation of Korea under Contract No. NRF-2022R1A2C1092335; National Science and Technology fund of Mongolia; Polish National Science Centre under Contract No. 2024/53/B/ST2/00975; STFC (United Kingdom); Swedish Research Council under Contract No. 2019.04595; U. S. Department of Energy under Contract No. DE-FG02-05ER41374



\bibliographystyle{JHEP}
\bibliography{references}

\appendix
\section{$K_i^{(\prime)}$ for $D\to\kslpipi$}
\label{sec:ki}
As part of the dataset used in this analysis overlaps with that in Ref.~\cite{BESIII:2025nsp}, only results obtained using the non-overlapping tag modes $D\to K^-e^+\nu_e$ and $D\to K\pi$ from Ref.~\cite{BESIII:2025nsp} are employed in the present work. By combining the $K_i^{(\prime)}$ values extracted from these modes, we obtain a set of $K_i^{(\prime)}$ parameters from the $\chi^2$ fit with improved precision compared to the final results that used all tag channels in Ref.~\cite{BESIII:2025nsp}. The resulting values are summarized in Table~\ref{tab:kspipiki_new}, and the corresponding correlation matrix is provided in Table~\ref{tab:ki_cor}.

\begin{table}[h]
	\centering
	\caption{The $K_i$ parameters for $D\to\kspipi$ and $K_i^{\prime}$ parameters for $D\to\klpipi$ given by $\chi^2$ fit with both statistical and systematic uncertainties included.}
	\begin{tabular}{lcc}
		\toprule
		Bins&$K_i$&$K_i^\prime$\\
		\midrule
-8&0.0292$\pm$0.0005 & 0.0346$\pm$0.0005\\
-7&0.0130$\pm$0.0004 & 0.0161$\pm$0.0005\\
-6&0.0143$\pm$0.0003 & 0.0128$\pm$0.0004\\
-5&0.0495$\pm$0.0007 & 0.0428$\pm$0.0004\\
-4&0.0165$\pm$0.0003 & 0.0141$\pm$0.0003\\
-3&0.0204$\pm$0.0004 & 0.0220$\pm$0.0004\\
-2&0.0183$\pm$0.0004 & 0.0228$\pm$0.0003\\
-1&0.0812$\pm$0.0009 & 0.0938$\pm$0.0009\\
1&0.1726$\pm$0.0017 & 0.1753$\pm$0.0016\\
2&0.0882$\pm$0.0009 & 0.0826$\pm$0.0008\\
3&0.0662$\pm$0.0008 & 0.0658$\pm$0.0007\\
4&0.0253$\pm$0.0004 & 0.0275$\pm$0.0004\\
5&0.0873$\pm$0.0009 & 0.0807$\pm$0.0008\\
6&0.0599$\pm$0.0007 & 0.0577$\pm$0.0006\\
7&0.1258$\pm$0.0013 & 0.1214$\pm$0.0013\\
8&0.1333$\pm$0.0015 & 0.1279$\pm$0.0017\\
		\bottomrule
	\end{tabular}
	\label{tab:kspipiki_new}
\end{table}

\newpage
\begin{table}[h]
	\centering
	\tiny
\caption{Correlation coefficients matrix (\%) for the fraction $K_i^{(\prime)}$.}
\begin{tabular}{lcccccccccccccccc}
	\toprule
	& $K_{-8}$ & $K_{-7}$ & $K_{-6}$ & $K_{-5}$ & $K_{-4}$ & $K_{-3}$ & $K_{-2}$ & $K_{-1}$ & $K_{1}$ & $K_{2}$ & $K_{3}$ & $K_{4}$ & $K_{5}$ & $K_{6}$ & $K_{7}$ & $K_{8}$ \\
    \midrule
$K_{-8}$ & 100      & 7.3      & 32.7     & 28.8     & 16.4     & 2.8      & 19.3     & 15.7     & 35.6    & 36.7    & 24.5    & 29.6    & 33      & 23.5    & 39.2    & 52.3    \\
$K_{-7}$ &          & 100      & 15.2     & 21.5     & -11.5    & -10      & -26.6    & 36.8     & 18.7    & 27.1    & 30.9    & 21.6    & 11.8    & 14.6    & 4.9     & -12.3   \\
$K_{-6}$ &          &          & 100      & 36.5     & 14.5     & -21.9    & 0.2      & 11.5     & 30.5    & 16.1    & -0.2    & 7.1     & 5.9     & 3.2     & 24.8    & 31.4    \\
$K_{-5}$ &          &          &          & 100      & 21.7     & -10.1    & 5.3      & 33       & 49.8    & 30.9    & 11      & 15.2    & 20.4    & 19.4    & 41.5    & 39.1    \\
$K_{-4}$ &          &          &          &          & 100      & 5.7      & 23.7     & 8.9      & 25.3    & 16      & 2.5     & 3.1     & 14.7    & 13.5    & 30.6    & 34      \\
$K_{-3}$ &          &          &          &          &          & 100      & 24.6     & 21.3     & 5.6     & 29.1    & 39.7    & 28.5    & 43.2    & 39.1    & 19.7    & 13.8    \\
$K_{-2}$ &          &          &          &          &          &          & 100      & 7.2      & 22.2    & 21.2    & 12.8    & 12.4    & 26.3    & 23.1    & 33.8    & 39.4    \\
$K_{-1}$ &          &          &          &          &          &          &          & 100      & 53.4    & 52.3    & 47      & 37.1    & 43.2    & 44.4    & 44.1    & 26.1    \\
$K_{1}$  &          &          &          &          &          &          &          &          & 100     & 52.6    & 30.4    & 31.5    & 40.6    & 39.4    & 63.6    & 57.4    \\
$K_{2}$  &          &          &          &          &          &          &          &          &         & 100     & 54.5    & 45.8    & 55.1    & 50.4    & 52.2    & 44.2    \\
$K_{3}$  &          &          &          &          &          &          &          &          &         &         & 100     & 48      & 53.8    & 48.9    & 33.2    & 21.9    \\
$K_{4}$  &          &          &          &          &          &          &          &          &         &         &         & 100     & 45.9    & 39.9    & 33.2    & 28.2    \\
$K_{5}$  &          &          &          &          &          &          &          &          &         &         &         &         & 100     & 52.4    & 47.3    & 42.7    \\
$K_{6}$  &          &          &          &          &          &          &          &          &         &         &         &         &         & 100     & 43.7    & 34.7    \\
$K_{7}$  &          &          &          &          &          &          &          &          &         &         &         &         &         &         & 100     & 62.1    \\
$K_{8}$  &          &          &          &          &          &          &          &          &         &         &         &         &         &         &         & 100     \\\bottomrule\toprule
\multicolumn{1}{c}{} & $K_{-8}^{\prime}$ & $K_{-7}^{\prime}$ & $K_{-6}^{\prime}$ & $K_{-5}^{\prime}$ & $K_{-4}^{\prime}$ & $K_{-3}^{\prime}$ & $K_{-2}^{\prime}$ & $K_{-1}^{\prime}$ & $K_{1}^{\prime}$ & $K_{2}^{\prime}$ & $K_{3}^{\prime}$ & $K_{4}^{\prime}$ & $K_{5}^{\prime}$ & $K_{6}^{\prime}$ & $K_{7}^{\prime}$ & $K_{8}^{\prime}$ \\\midrule
$K_{-8}^{\prime}$    & 100               & 12.5              & 20.4              & 35.7              & 5.5               & 32.5              & 41.7              & 54.1              & 29.8             & 50.8             & 33.1             & 46.7             & 60.7             & 44               & 22.4             & 42               \\
$K_{-7}^{\prime}$    &                   & 100               & -10.2             & 13.6              & -9.2              & 6.4               & -8.1              & 36.8              & 32.8             & 16.8             & 10.2             & 51.8             & 12.7             & 36.1             & 10.1             & -11              \\
$K_{-6}^{\prime}$    &                   &                   & 100               & -4.3              & 24.8              & -9.3              & 33                & 19.9              & 14.4             & 12.4             & 43               & -16              & 2.9              & 4.4              & 29.8             & 17.1             \\
$K_{-5}^{\prime}$    &                   &                   &                   & 100               & 12.7              & 32.6              & 37.5              & 53.3              & 51.4             & 56.4             & 41.3             & 31.6             & 47.3             & 47.3             & 38.8             & 37.3             \\
$K_{-4}^{\prime}$    &                   &                   &                   &                   & 100               & -10.1             & 18                & 11.4              & 29.7             & 29.2             & 34.6             & -1.4             & 16.3             & 15.5             & 31               & 25.1             \\
$K_{-3}^{\prime}$    &                   &                   &                   &                   &                   & 100               & 35.7              & 41.3              & 15.1             & 19.9             & -2               & 32.1             & 42.1             & 40.7             & 16.2             & 21.2             \\
$K_{-2}^{\prime}$    &                   &                   &                   &                   &                   &                   & 100               & 56.3              & 34.6             & 45.7             & 46.4             & 9.1              & 35.6             & 41.2             & 36               & 44.3             \\
$K_{-1}^{\prime}$    &                   &                   &                   &                   &                   &                   &                   & 100               & 58.6             & 61.6             & 56               & 40.9             & 48.7             & 60.7             & 44.5             & 37.9             \\
$K_{1}^{\prime}$     &                   &                   &                   &                   &                   &                   &                   &                   & 100              & 61.1             & 62.3             & 37.5             & 52.3             & 41.2             & 80.9             & 13.8             \\
$K_{2}^{\prime}$     &                   &                   &                   &                   &                   &                   &                   &                   &                  & 100              & 61.9             & 38.2             & 55.5             & 61               & 42.8             & 62.8             \\
$K_{3}^{\prime}$     &                   &                   &                   &                   &                   &                   &                   &                   &                  &                  & 100              & 6.3              & 27.1             & 34.6             & 54.1             & 38.7             \\
$K_{4}^{\prime}$     &                   &                   &                   &                   &                   &                   &                   &                   &                  &                  &                  & 100              & 60.3             & 50.8             & 21.4             & 18.2             \\
$K_{5}^{\prime}$     &                   &                   &                   &                   &                   &                   &                   &                   &                  &                  &                  &                  & 100              & 51.2             & 48.3             & 36               \\
$K_{6}^{\prime}$     &                   &                   &                   &                   &                   &                   &                   &                   &                  &                  &                  &                  &                  & 100              & 23.7             & 62.3             \\
$K_{7}^{\prime}$     &                   &                   &                   &                   &                   &                   &                   &                   &                  &                  &                  &                  &                  &                  & 100              & 0.4              \\
$K_{8}^{\prime}$     &                   &                   &                   &                   &                   &                   &                   &                   &                  &                  &                  &                  &                  &                  &                  & 100             \\\bottomrule\toprule
\multicolumn{1}{c}{} & $K_{-8}^{\prime}$ & $K_{-7}^{\prime}$ & $K_{-6}^{\prime}$ & $K_{-5}^{\prime}$ & $K_{-4}^{\prime}$ & $K_{-3}^{\prime}$ & $K_{-2}^{\prime}$ & $K_{-1}^{\prime}$ & $K_{1}^{\prime}$ & $K_{2}^{\prime}$ & $K_{3}^{\prime}$ & $K_{4}^{\prime}$ & $K_{5}^{\prime}$ & $K_{6}^{\prime}$ & $K_{7}^{\prime}$ & $K_{8}^{\prime}$ \\\midrule
$K_{-8}$             & -8.9              & -11.5             & 2.1               & 7.4               & 7.1               & 10                & 3.5               & -3.8              & -4.7             & -4.8             & -4.1             & -6.8             & 4.2              & 1.9              & 2.3              & 0.3              \\
$K_{-7}$             & -9.7              & 3.6               & 3.2               & -11.3             & 23.2              & -7.3              & -5.3              & -10.1             & -2.3             & -1               & -6.5             & 5.6              & 2.1              & 9.1              & -2.3             & 7.2              \\
$K_{-6}$             & -8.3              & -2.2              & 0.6               & -0.6              & 8.6               & 1.3               & -1.4              & -5.6              & -2.1             & -2               & -4.3             & -0.2             & 1.9              & 4.1              & -0.1             & 3                \\
$K_{-5}$             & -6.9              & 7.5               & 0.1               & -1.4              & 6.2               & 2.1               & -2.3              & -2.1              & -1.7             & -2.9             & -3.7             & 4.2              & 0.5              & 6.5              & -2.6             & 0.2              \\
$K_{-4}$             & -7                & 4.9               & -6                & -1.1              & -5.9              & -5.3              & -9.8              & -6.4              & -3.9             & -4.8             & -3.8             & -0.8             & -5.7             & -4.3             & -6.2             & -5.8             \\
$K_{-3}$             & 10.3              & 9.8               & -0.2              & 6.8               & -13.3             & 5.1               & 0.9               & 9.4               & -2.9             & -2.1             & 3.6              & 2.2              & -1.4             & -2.7             & -5.5             & -9               \\
$K_{-2}$             & -3.4              & -5.3              & 1.8               & 2.7               & -8.1              & -0.8              & -2.4              & -1.9              & -5.2             & -6.7             & 1.6              & -8.8             & -5.8             & -9.5             & -2.7             & -8               \\
$K_{-1}$             & -6                & 1.8               & -0.1              & -2.6              & 9                 & 0                 & -2                & -3.9              & -4.4             & -4.6             & -2.1             & 0.2              & -0.6             & 1                & -5.2             & -2.9             \\
$K_{1}$              & -9.6              & -3.8              & -0.9              & 2.3               & 9.2               & 1.1               & -1.3              & -5                & -2.3             & -3.6             & -0.3             & -3.2             & 0.6              & 0.1              & -1.6             & -2               \\
$K_{2}$              & -6.9              & -1.3              & -2.6              & 2.3               & 7.1               & 3.5               & -2.2              & -4.2              & -6.3             & -5.7             & -4.6             & -1.4             & 1.1              & 0.9              & -5.2             & -3.8             \\
$K_{3}$              & -2.9              & -5.2              & -1.6              & 2.4               & 7.2               & 2.8               & 0.7               & -2                & -6.1             & -4               & -2.4             & -4.1             & 0.7              & -1.6             & -4.7             & -3.5             \\
$K_{4}$              & 0.5               & -6.6              & 4.7               & 5.5               & 7.2               & 7.2               & 7.2               & 3.3               & -2.8             & -1.7             & 2.1              & -4.8             & 3.2              & 0.7              & 0.6              & -1.4             \\
$K_{5}$              & 0.9               & 2.8               & -1.4              & 8                 & -1.4              & 9.5               & 2.9               & 4.8               & -3.8             & -3.4             & -0.5             & -0.2             & 1.7              & 1.8              & -3.4             & -5.4             \\
$K_{6}$              & -0.3              & 5.4               & 0.3               & 3.4               & 0.1               & 4                 & -0.3              & 2.2               & -3.6             & -3.8             & 0                & 1                & -0.1             & 0.8              & -4.4             & -5.4             \\
$K_{7}$              & -6                & 0.6               & 1.8               & 3.6               & 3.4               & -0.3              & -2.1              & -2.2              & -3.2             & -4.2             & 2                & -3.4             & -1.4             & -2.2             & -3               & -5.4             \\
$K_{8}$              & -7.1              & -6.1              & -3.8              & 10.8              & -2.3              & 8.4               & 0.2               & -1.4              & -3.3             & -4.6             & -1.1             & -5.8             & 1                & -2               & -0.6             & -5.2             \\\bottomrule
	\end{tabular}
\label{tab:ki_cor}
\end{table}

\section{$D\to \kmthreepi$ binned signal yields}
The signal yields of $C\!P$-tagged and Flavour-tagged events in the binned $D\to \kmthreepi$ phase space, obtained with the same method in Section~\ref{sec:selection} and corrected by Monte Carlo-determined efficiencies, are shown in Table~\ref{tab:binnedyields}.

\begin{table*}[ht]
\begin{center}
\setlength\tabcolsep{6pt}
\caption{Efficiency-corrected signal yields for $C\!P$-tagged and flavour-tagged $\kmthreepi$ events in each $\kmthreepi$ bin, where the uncertainties are statistical only.}
\begin{tabular}{cccccc}
\toprule
Mode                             & Bin & 1                   & 2                   & 3                   & 4                   \\ \midrule
$\kmppi$                         &     & $278\pm56$      & $75\pm31$       & $59\pm30$       & $305\pm62$      \\
$\kmppipio$                      &     & $757\pm89$      & $448\pm64$      & $413\pm68$      & $683\pm88$      \\
\multirow{4}{*}{$\kmppipipi$}    & 1   & $49\pm21$       & $94\pm29$       & $56\pm24$       & $203\pm41$      \\
                                 & 2   &                     & $47\pm19$       & $66\pm25$       & $133\pm32$      \\
                                 & 3   &                     &                     & $26\pm17$       & $76\pm27$       \\
                                 & 4   &                     &                     &                     & $90\pm32$       \\
$\kpmpi$                         &     & $43412\pm413$   & $37533\pm370$   & $39705\pm387$   & $49996\pm447$   \\
$\kpmpipio$                      &     & $159060\pm1131$ & $139985\pm1033$ & $144937\pm1064$ & $187033\pm1249$ \\
\multirow{4}{*}{$\kpmpipipi$}    & 1   & $10495\pm276$   & $19184\pm360$   & $19506\pm367$   & $24946\pm436$   \\
                                 & 2   &                     & $7890\pm223$    & $17280\pm336$   & $22406\pm399$   \\
                                 & 3   &                     &                     & $8815\pm248$    & $22615\pm406$   \\
                                 & 4   &                     &                     &                     & $15013\pm344$   \\
$\kspio$                         &     & $8407\pm233$    & $7657\pm212$    & $7094\pm209$    & $10324\pm260$   \\
$\ks\eta(\gamma\gamma)$          &     & $1437\pm108$    & $1172\pm95$     & $1030\pm90$     & $1745\pm124$    \\
$\ks\eta(\pi\pi\pi^0)$           &     & $800\pm108$     & $813\pm109$     & $680\pm101$     & $847\pm134$     \\
$\ks\eta^{\prime}(\eta\pi\pi)$   &     & $966\pm131$     & $1075\pm132$    & $919\pm124$     & $1452\pm162$    \\
$\ks\eta^{\prime}(\pi\pi\gamma)$ &     & $2146\pm169$    & $1963\pm160$    & $1908\pm158$    & $2471\pm185$    \\
$\ksomega$                       &     & $6677\pm409$    & $6187\pm370$    & $6414\pm381$    & $9079\pm485$    \\
$\ks\phi$                        &     & $773\pm172$     & $773\pm156$     & $1205\pm172$    & $1376\pm191$    \\
$\klpiopio$                      &     & $9382\pm500$    & $7814\pm457$    & $8195\pm463$    & $11342\pm569$   \\
$\klomega$                       &     & $10386\pm518$   & $10134\pm423$   & $9535\pm467$    & $12750\pm549$   \\
$\kk$                            &     & $4879\pm136$    & $4339\pm126$    & $4645\pm129$    & $5314\pm143$    \\
$\pipipio$                       &     & $15808\pm339$   & $14406\pm314$   & $15059\pm328$   & $17749\pm371$   \\
$\pipi$                          &     & $1894\pm82$     & $1454\pm71$     & $1550\pm78$     & $2082\pm89$     \\
$\kspiopio$                      &     & $7025\pm359$    & $7194\pm355$    & $7350\pm361$    & $8005\pm395$    \\
$\klpio$                         &     & $10031\pm442$   & $9543\pm428$    & $9593\pm438$    & $10736\pm488$   \\ \bottomrule
\end{tabular}
\label{tab:binnedyields}
\end{center}
\end{table*}

\section{Systematic uncertainties and migration matrices of binned analysis}
The systematic uncertainties are included in the fit for determining the binned strong-phase parameters in the same manner as the global fit. The systematic uncertainties are reported in Table~\ref{tab:inputsys_bin}.

The systematic uncertainty due to $K^\pm,\pi^\pm$ tracking and identification is assigned independently for $D\to K^-\pi^+$, $D\to K^-\pi^+\pi^0$, and $D\to K^-\pi^+\pi^+\pi^-$, but is fully correlated across the four bins of phase space. The tracking and identification uncertainties are also fully correlated between the different tag modes and across the four bins of phase space. The uncertainties due to $\pi^0$ reconstruction, impact of resonance modelling on efficiency, and fit method for signal yields are also completely correlated across the tag modes. The phase space region populated by the $D \to K_S^0K^-\pi^+$ decay is excluded in the measurement of the strong-phase parameters in the binned analysis and therefore does not contribute to the systematic uncertainty. The central values and uncertainties of the external parameters used at the time of the analysis can be found in Table~\ref{tab:inputs}. All systematic uncertainties arising from different sources are treated as uncorrelated.

\begin{table}[htbp]
\caption{Systematic uncertainties contributing to the determination of the binned hadronic parameters.}
\label{tab:inputsys_bin}
\begin{center}\smaller
\begin{tabular}{ll}
\toprule
 Systematics & Sizes\\
\midrule
$K^\pm, \pi^\pm$ tracking and identification &0.26\% for each $K^\pm$ and 0.15\% for each $\pi^\pm$   \\
$\pi^0$ reconstruction &1\%, only contributing in $D\to K^-\pi^+\pi^0$\\
Impact of resonance modelling on efficiency &2\% for DCS $D\to K^-\pi^+\pi^0$\\
$D \to K_S^0K^-\pi^+$ background &Negligible as excluded by the binning scheme\\
Fit method for signal yields &1.4\% for $D\to K^-\pi^+\pi^+\pi^-$ and 2.0\% for $D\to K^-\pi^+\pi^0$\\	
External parameters &See Table~\ref{tab:inputs}\\
\bottomrule
\end{tabular}
\end{center}
\end{table}

The migration effects among $D\to\kmpipipi$ phase-space bins for various tag modes are determined with dedicated Monte Carlo simulation samples. The migration matrices are listed in Tables~\ref{tab:binmigration1}-\ref{tab:binmigration6}, and show that $\gtrsim\;$90$\%$ of events are reconstructed correctly in the corresponding bins for flavour and $C\!P$ tags; $80-95\%$ events are reconstructed correctly in the corresponding bins for the $D\to\kspipi$ tag and $70-95\%$ events are reconstructed correctly in the corresponding bins for the $D\to\klpipi$ tag. The wrongly assigned events lie mostly in the neighbouring bins. These effects are taken into account when calculating the efficiency-corrected signal yields and induce correlations among the four $D\to\kmthreepi$ bins, which must be taken into consideration when determining the covariance matrices.

\begin{sidewaystable*}[!ht]
\caption{Migration effects ($\%$) in the flavour and $C\!P$ tag modes. The columns and rows are the truth and reconstructed bins, respectively.}\label{tab:binmigration1}
\begin{center}
\setlength\tabcolsep{4pt}
\begin{tabular}{c cccc| cccc |cccc |cccc}
\toprule
    & \multicolumn{4}{c|}{$\kmpi$}               & \multicolumn{4}{c|}{$\kmpipio$}             & \multicolumn{4}{c|}{$\kk$}                 & \multicolumn{4}{c}{$\pipi$}                 \\
Bin & 1         & 2         & 3        & 4       & 1          & 2         & 3        & 4       & 1         & 2        & 3        & 4        & 1          & 2        & 3        & 4        \\
1   & 96.17     & 1.86      & 0.16     & 1.29    & 96.08      & 1.91      & 0.15     & 1.31    & 96.55     & 1.95     & 0.07     & 0.78     & 96.64      & 1.91     & 0.08     & 0.74     \\
2   & 2.99      & 93.87     & 2.37     & 0.20    & 2.99       & 93.85     & 2.34     & 0.20    & 2.48      & 94.75    & 2.07     & 0.05     & 2.43       & 94.74    & 2.09     & 0.07     \\
3   & 0.26      & 3.52      & 93.22    & 2.59    & 0.24       & 3.58      & 93.15    & 2.58    & 0.05      & 3.16     & 93.81    & 2.32     & 0.08       & 3.02     & 94.00    & 2.24     \\
4   & 0.95      & 0.19      & 3.32     & 94.91   & 0.95       & 0.18      & 3.33     & 94.84   & 0.51      & 0.05     & 2.68     & 96.23    & 0.49       & 0.07     & 2.69     & 96.19    \\ \midrule
    & \multicolumn{4}{c|}{$\pipipio$}            & \multicolumn{4}{c|}{$\kspiopio$}            & \multicolumn{4}{c|}{$\klpio$}              & \multicolumn{4}{c}{$\klomega$}              \\
Bin & 1         & 2         & 3        & 4       & 1          & 2         & 3        & 4       & 1         & 2        & 3        & 4        & 1          & 2        & 3        & 4        \\
\bottomrule
\end{tabular}
\end{center}
\end{sidewaystable*}

\begin{table*}[!ht]
\caption{Migration effects ($\%$)  for the $D\to\kmthreepi$ events tagged with $D\to\kmthreepi$. The columns and rows are the truth and reconstructed bins, respectively.  Values below 0.005 are recorded as -.}\label{tab:binmigration2}
\begin{center}
\setlength\tabcolsep{6pt}
\begin{tabular}{c| cccccccccc}
\toprule
Bin & 11    & 21    & 22    & 31    & 32    & 33    & 41    & 42    & 43    & 44    \\ \midrule
11  & 90.56 & 3.86  & 0.11  & 0.61  & 0.08  & 0.05  & 2.33  & 0.29  & 0.19  & 0.13  \\
21  & 2.68  & 89.47 & 1.55  & 2.44  & 0.32  & 0.04  & 0.31  & 1.70  & 0.15  & 0.09  \\
22  & 0.23  & 6.13  & 85.38 & 0.23  & 5.27  & -     & 0.27  & 0.36  & 0.09  & 0.14  \\
31  & 0.14  & 2.60  & 0.07  & 90.05 & 1.48  & 0.15  & 2.48  & 0.37  & 1.19  & 0.06  \\
32  & 0.13  & 0.50  & 2.63  & 2.77  & 87.16 & 2.27  & 0.40  & 2.52  & 0.31  & 0.05  \\
33  & 0.08  & 0.31  & 0.31  & 0.58  & 5.72  & 86.03 & 0.12  & 0.47  & 4.83  & 0.16  \\
41  & 0.62  & 0.21  & 0.09  & 3.01  & 0.16  & 0.02  & 91.13 & 2.26  & 0.35  & 0.89  \\
42  & 0.09  & 0.93  & 0.06  & 0.24  & 2.33  & 0.18  & 3.07  & 89.44 & 2.15  & 0.14  \\
43  & 0.07  & 0.25  & -     & 0.97  & 0.41  & 2.32  & 0.45  & 3.50  & 88.71 & 2.24  \\
44  & 0.19  & 0.24  & 0.12  & 0.33  & 0.14  & 0.14  & 1.83  & 0.40  & 5.71  & 89.45\\
\bottomrule
\end{tabular}
\end{center}
\end{table*}

\begin{sidewaystable*}[!ht]
\begin{center}\tiny
\setlength\tabcolsep{2pt}
\caption{Migration effects ($\%$) for the $D\to\kmthreepi$ events tagged with $D\to\kspipi$ reconstructed in the $D\to\kspipi$ bins from 1 to 8. The columns and rows are the truth and reconstructed bins respectively. Values below 0.005 are recorded as -.}
\label{tab:binmigration3}
\begin{tabular}{cc cccc cccc cccc cccc cccc cccc cccc cccc}
\toprule
& &\multicolumn{4}{c}{$1$} &\multicolumn{4}{c}{$2$} &\multicolumn{4}{c}{$3$} &\multicolumn{4}{c}{$4$} &\multicolumn{4}{c}{$5$} &\multicolumn{4}{c}{$6$} &\multicolumn{4}{c}{$7$} &\multicolumn{4}{c}{$8$}\\
& &1 &2 &3 &4 &1 &2 &3 &4 &1 &2 &3 &4 &1 &2 &3 &4 &1 &2 &3 &4 &1 &2 &3 &4 &1 &2 &3 &4 &1 &2 &3 &4\\
\multirow{4}{*}{$-8$} & 1 & - & - & - & - & - & - & - & - & - & - & - & - & - & - & - & - & - & - & - & - & - & - & - & - & - & - & - & - & 0.11 & - & - & - \\
 & 2 & - & 0.11 & - & - & - & - & - & - & - & - & - & - & - & - & - & - & - & - & - & - & - & - & - & - & - & - & - & - & - & 0.11 & - & - \\
 & 3 & - & 0.11 & 0.11 & - & - & - & - & - & - & - & - & - & - & - & - & - & - & - & - & - & - & - & - & - & - & - & - & - & - & - & - & - \\
 & 4 & - & - & - & - & - & - & - & - & - & - & - & 0.11 & - & - & - & - & - & - & - & - & - & - & - & - & - & - & - & - & - & - & - & - \\
\multirow{4}{*}{$-7$} & 1 & 0.22 & - & - & - & - & - & 0.22 & - & - & - & - & - & - & - & - & - & - & - & - & - & - & - & - & - & - & - & - & - & 0.44 & - & - & - \\
 & 2 & - & - & 0.24 & - & - & - & - & - & - & - & - & - & - & - & - & - & - & - & - & - & - & - & - & - & - & - & - & - & - & - & - & - \\
\bottomrule
\end{tabular}
\end{center}
\end{sidewaystable*}

\begin{sidewaystable*}[!ht]
\begin{center}\tiny
\caption{Migration effects ($\%$)  for the $D\to\kmthreepi$ events tagged with $D\to\kspipi$ reconstructed in the $D\to\kspipi$ bins from $-$1 to $-$8. The columns and rows are the truth and reconstructed bins respectively. Values below 0.005 are recorded as -.}
\label{tab:binmigration4}
\setlength\tabcolsep{2pt}
\begin{tabular}{cc cccc cccc cccc cccc cccc cccc cccc cccc}
\toprule
 & &\multicolumn{4}{c}{$-8$} &\multicolumn{4}{c}{$-7$} &\multicolumn{4}{c}{$-6$} &\multicolumn{4}{c}{$-5$} &\multicolumn{4}{c}{$-4$} &\multicolumn{4}{c}{$-3$} &\multicolumn{4}{c}{$-2$} &\multicolumn{4}{c}{$-1$}\\
& &1 &2 &3 &4 &1 &2 &3 &4 &1 &2 &3 &4 &1 &2 &3 &4 &1 &2 &3 &4 &1 &2 &3 &4 &1 &2 &3 &4 &1 &2 &3 &4\\
\multirow{4}{*}{$-8$} & 1 & 90.93 & 1.33 & - & 0.33 & 1.99 & - & - & - & - & - & - & - & 0.22 & - & - & - & - & - & - & - & - & - & - & - & - & - & - & - & 4.20 & 0.33 & - & 0.11 \\
 & 2 & 2.06 & 88.42 & 1.62 & - & 0.11 & 1.41 & - & - & - & - & - & - & - & 0.32 & - & - & - & - & - & - & - & 0.11 & - & - & - & 0.22 & - & - & 0.22 & 4.98 & - & - \\
 & 3 & 0.11 & 4.03 & 87.46 & 1.23 & - & - & 0.90 & - & - & - & - & - & 0.22 & - & - & - & - & - & - & - & - & - & - & - & - & - & 0.11 & - & - & 0.22 & 4.26 & 0.11 \\
 & 4 & 0.84 & - & 2.22 & 89.65 & - & - & 0.11 & 1.80 & - & - & - & 0.11 & - & - & - & - & - & - & - & 0.11 & - & - & - & - & - & - & - & - & - & - & 0.21 & 4.54 \\
\multirow{4}{*}{$-7$} & 1 & 3.09 & - & - & - & 89.18 & 1.55 & - & 0.66 & 3.53 & 0.22 & - & - & 0.22 & - & - & - & - & - & - & - & 0.22 & - & - & - & 0.22 & - & - & - & - & - & - & - \\
 & 2 & 0.24 & 3.79 & - & - & 2.84 & 87.44 & 1.42 & - & - & 2.13 & - & - & - & 0.47 & - & - & - & - & - & - & - & - & - & - & - & - & - & - & 0.24 & 0.47 & 0.24 & - \\
\bottomrule
\end{tabular}
\end{center}
\end{sidewaystable*}

\begin{sidewaystable*}[!ht]
\begin{center}\tiny
\setlength\tabcolsep{2pt}
\caption{Migration effects ($\%$) for the $D\to\kmthreepi$ events tagged with $D\to\klpipi$ reconstructed in the $D\to\klpipi$ bins from 1 to 8. The columns and rows are the truth and reconstructed bins respectively. Values below 0.005 are recorded as -.}
\label{tab:binmigration5}
\begin{tabular}{cc cccc cccc cccc cccc cccc cccc cccc cccc}
\toprule
& &\multicolumn{4}{c}{$1$} &\multicolumn{4}{c}{$2$} &\multicolumn{4}{c}{$3$} &\multicolumn{4}{c}{$4$} &\multicolumn{4}{c}{$5$} &\multicolumn{4}{c}{$6$} &\multicolumn{4}{c}{$7$} &\multicolumn{4}{c}{$8$}\\
& &1 &2 &3 &4 &1 &2 &3 &4 &1 &2 &3 &4 &1 &2 &3 &4 &1 &2 &3 &4 &1 &2 &3 &4 &1 &2 &3 &4 &1 &2 &3 &4\\
\multirow{4}{*}{$-8$} & 1 & 0.04 & - & - & - & - & - & - & - & - & - & - & - & - & - & - & - & - & - & - & - & - & - & - & - & 0.07 & - & - & - & 0.04 & - & - & 0.04 \\
 & 2 & - & - & - & - & - & - & - & - & - & - & - & - & - & - & - & - & - & - & - & - & - & 0.04 & - & - & - & - & - & - & - & 0.04 & - & - \\
 & 3 & - & - & 0.12 & - & - & - & - & - & - & - & - & - & - & 0.04 & - & - & - & - & - & - & - & - & - & - & 0.04 & - & - & - & - & - & 0.04 & - \\
 & 4 & - & - & - & 0.10 & - & - & - & - & - & - & - & - & - & - & - & - & - & - & - & - & - & - & - & - & - & - & - & 0.03 & - & - & - & - \\
\multirow{4}{*}{$-7$} & 1 & 0.16 & - & - & - & - & - & - & - & - & - & - & - & - & - & - & - & - & - & - & - & - & - & - & - & - & - & - & - & 0.08 & - & - & - \\
 & 2 & - & 0.08 & - & - & - & - & - & - & 0.08 & - & - & - & - & - & - & - & - & - & - & - & - & - & - & - & - & 0.08 & - & - & - & - & - & - \\
\bottomrule
\end{tabular}
\end{center}
\end{sidewaystable*}

\begin{sidewaystable*}[!ht]
\begin{center}\tiny
\caption{Migration effects ($\%$)  for the $D\to\kmthreepi$ events tagged with $D\to\klpipi$ reconstructed in the $D\to\klpipi$ bins from $-$1 to $-$8. The columns and rows are the truth and reconstructed bins respectively. Values below 0.005 are recorded as -.}
\label{tab:binmigration6}
\setlength\tabcolsep{2pt}
\begin{tabular}{cc cccc cccc cccc cccc cccc cccc cccc cccc}
\toprule
 & &\multicolumn{4}{c}{$-8$} &\multicolumn{4}{c}{$-7$} &\multicolumn{4}{c}{$-6$} &\multicolumn{4}{c}{$-5$} &\multicolumn{4}{c}{$-4$} &\multicolumn{4}{c}{$-3$} &\multicolumn{4}{c}{$-2$} &\multicolumn{4}{c}{$-1$}\\
& &1 &2 &3 &4 &1 &2 &3 &4 &1 &2 &3 &4 &1 &2 &3 &4 &1 &2 &3 &4 &1 &2 &3 &4 &1 &2 &3 &4 &1 &2 &3 &4\\
\multirow{4}{*}{$-8$} & 1 & 81.98 & 1.42 & - & 0.50 & 3.22 & 0.11 & 0.04 & 0.04 & 0.18 & - & - & 0.04 & 0.07 & 0.04 & - & - & 0.04 & - & - & - & 0.14 & - & - & - & 0.28 & - & - & - & 10.80 & 0.11 & - & 0.11 \\
 & 2 & 2.23 & 79.71 & 1.81 & 0.13 & 0.08 & 3.36 & 0.04 & - & - & 0.08 & - & - & - & 0.08 & 0.04 & - & - & 0.08 & - & - & 0.08 & 0.25 & 0.04 & - & - & 0.63 & - & - & 0.29 & 9.96 & 0.21 & - \\
 & 3 & - & 2.64 & 80.17 & 1.77 & - & 0.04 & 3.31 & 0.08 & - & - & 0.08 & 0.04 & - & - & 0.12 & - & - & - & - & - & - & - & 0.20 & 0.04 & - & 0.08 & 0.47 & - & 0.04 & 0.28 & 9.31 & 0.20 \\
 & 4 & 0.52 & 0.06 & 2.04 & 82.14 & 0.03 & - & 0.03 & 3.82 & - & - & - & 0.19 & - & 0.03 & - & 0.16 & - & - & - & 0.06 & - & 0.06 & - & 0.06 & 0.03 & 0.03 & 0.03 & 0.55 & 0.03 & - & 0.45 & 8.99 \\
\multirow{4}{*}{$-7$} & 1 & 9.75 & 0.47 & - & 0.08 & 79.01 & 0.86 & 0.08 & 0.71 & 4.40 & 0.08 & - & 0.16 & 0.31 & - & - & 0.08 & 0.08 & - & - & - & 0.16 & - & - & - & 0.63 & - & - & - & 2.44 & - & - & - \\
 & 2 & 0.42 & 7.02 & 0.17 & - & 1.35 & 78.36 & 1.78 & 0.17 & 0.17 & 4.48 & 0.08 & - & 0.17 & 0.34 & - & - & 0.08 & 0.34 & - & 0.08 & - & 0.51 & - & - & 0.08 & 0.42 & 0.08 & - & 0.17 & 3.21 & 0.08 & - \\
\bottomrule
\end{tabular}
\end{center}
\end{sidewaystable*}

\end{document}